\documentclass[letterpaper]{article}
\pdfoutput=1
\usepackage{jheppub}
\usepackage{amsmath}
\usepackage{graphicx}% Include figure files
\usepackage{feynmf}
\usepackage{array}
\usepackage{slashed}
\usepackage{appendix}
\usepackage{multirow}
\usepackage{xcolor}
\usepackage{aas_macros} 

\usepackage{silence}
\newcommand{\bea}{\begin{eqnarray}}
\newcommand{\eea}{\end{eqnarray}}

\newcommand{\alphaem}{\alpha_{\rm em}}
\newcommand{\anomaly}{\mathcal{A}}
\title{A\;CMB\;Millikan\;Experiment\;with\;Cosmic\;Axiverse\;Strings}

\date{\today}
\author[a]{Prateek Agrawal,}
\author[b]{Anson Hook,}
\author[c]{and Junwu Huang}

\affiliation[a]{Jefferson Physical Laboratory, Harvard University,
17 Oxford Street, Cambridge, MA 02138, USA}
\affiliation[b]{Maryland Center for Fundamental Physics, University of Maryland, College Park, MD 20742, USA}
\affiliation[c]{Perimeter Institute for Theoretical Physics, 31 Caroline St.~N., Waterloo, Ontario N2L 2Y5, Canada}
\emailAdd{prateekagrawal@fas.harvard.edu}
\emailAdd{hook@umd.edu}
\emailAdd{jhuang@perimeterinstitute.ca}

\abstract{ We study axion strings of hyperlight axions coupled to photons. Hyperlight axions -- axions lighter than Hubble at recombination -- are a generic prediction of the string axiverse. These axions strings produce a distinct quantized polarization rotation of CMB photons which is $\mathcal{O}(\alphaem)$.  As the
  CMB light passes many strings, this polarization rotation converts
  E-modes to B-modes and adds up like a random walk. Using numerical
  simulations we show that the expected size
  of the final result is well within the reach of current and future
  CMB experiments through the measurement of correlations of CMB
  B-modes with E- and T-modes.  The quantized polarization rotation
  angle is topological in nature and can be seen as a geometric phase.
  Its value depends only on the anomaly coefficient and is independent
  of other details such as the axion decay constant.  Measurement of
  the anomaly coefficient by measuring this rotation will provide
  information about the UV theory, such as the quantization of
  electric charge and the value of the fundamental unit of charge.
  The presence of axion strings in the universe relies only on a phase
  transition in the early universe after inflation, after which the
  string network rapidly approaches an attractor scaling solution.  If
  there are additional stable topological objects such as domain
  walls, axions as heavy as $10^{-15}$ eV would be accessible.  The
  existence of these strings could also be probed by measuring the
relative polarization rotation angle between different images in
gravitationally lensed quasar systems.  }

\preprint{\today}

\begin{document}

\maketitle

\section{Introduction}

The ultraviolet structure of the standard model is one of the most
fundamental questions in particle physics.  In general, it is a
daunting task to unearth high-energy properties of the theory from
low-energy observations. However, there is a unique type of
interaction that contains universal information about the fundamental
theory unpolluted by intervening physics.  This special type of
interaction is a topological interaction~\cite{Dirac:1931kp}.  In this
paper, we consider the topological coupling of a compact
pseudo-scalar---an axion---with photons, 
\begin{align}
  \mathcal{L}
  &=
  \frac{\anomaly \alphaem}{4\pi} \frac{a}{f} F \widetilde F
  \label{eq: best interaction}
\end{align}
where $2 \pi f$ is the periodicity of the axion $a$, 
$\alphaem$ is the fine structure constant, and $\anomaly$ is
the anomaly coefficient for the mixed anomaly between electromagnetism
and the continuous shift symmetry of the axion.
Anomaly matching fixes the form of the interaction at any
scale~\cite{tHooft:1979rat}, such that the
rational number $\anomaly$ is not renormalized
and intervening physics between the high energy
scale and the low energy scale cannot hide fundamental physics from
us.  As an example, minimal Grand Unified Theories (GUTs) make the
concrete prediction that $\anomaly$ is a multiple of $4/3$.

In order to fully utilize this coupling, one requires topological
objects such as axion strings rather than particles.  One reason for
this is that the periodicity of the axion ($2 \pi f$) does change from
the ultraviolet (UV) to the infrared (IR) due to wavefunction
renormalization. Particle interactions depend on $f$ and are also
affected by kinetic and mass mixing effects. This makes it extremely
challenging to uncover information about the UV physics, and at the
minimum would require the discovery of multiple interactions
(see~\cite{diCortona:2015ldu} and references within).  However,
compact fields such as the axion naturally admit string-like
solutions, where $a/f$ winds from $0$ to $2\pi$ as one goes around the
string. This non-trivial winding makes an infinitely long string
topologically stable, since a string solution cannot be continuously deformed
into a solution where $a/f = 0$ everywhere. The topological winding is
also independent of the value of $f$ or mixing effects. This implies
that string
observables can give us direct information about the UV physics!
For example, measurement of $\anomaly \sim Q^2$ can provide us
information about the fundamental unit of electric charge.  Thus,
measurements of
axion strings are a form of the classic Millikan
experiment~\cite{1913PhRv....2..109M}.  Fortunately, there is a
compelling possibility that the universe may set up such a Millikan
experiment for us in cosmology.

The first ingredient, hyperlight axions\footnote{By hyperlight
  axions, we mean axions with masses smaller than the Hubble parameter
  at CMB, $m_a \lesssim 10^{-27}$ eV. Such axions arise in string theory, but are too light to be
the QCD axion or the dark matter candidate.} with a photon coupling, are ubiquitous in string
theory
compactifications~\cite{axion1,axion2,axion3,
Svrcek:2006yi,Arvanitaki:2009fg,Demirtas:2018akl}.
Axions have a continuous shift symmetry that protects them from
getting a mass. This shift symmetry is approximate and is violated by
instantons, which typically results in exponentially
small masses for the axion. 
Axions also quite generically
have interactions of the form shown in equation~\eqref{eq: best interaction}
which descend naturally from compactifications of
higher dimensional theories.  This combination has motivated the
picture of the string axiverse~\cite{Arvanitaki:2009fg} with a
plethora of low-energy axions potentially accessible to experiments
today through their photon
interactions~\cite{Sikivie:1983ip,Krauss:1985ub,
Sikivie:1985yu,Asztalos:2009yp,Kahn:2016aff,Brubaker:2016ktl,TheMADMAXWorkingGroup:2016hpc,
Baryakhtar:2018doz} (see~\cite{Graham:2013gfa,Irastorza:2018dyq} for recent reviews).
One of the most compelling candidates is the QCD axion, which
couples to gluons and solves the strong CP problem of the
Standard Model~\cite{axion1,axion2,axion3}. This has sparked a rich
experimental program searching for the QCD
axion~\cite{Sikivie:1983ip,Krauss:1985ub,
Sikivie:1985yu,Asztalos:2009yp,Kahn:2016aff,
Brubaker:2016ktl,Baryakhtar:2018doz,Arvanitaki:2014dfa}.  Axion-like
particles are axions that do not solve the strong CP problem.  Like
the QCD axion, there has also been a huge effort to discover these
particles~\cite{Sikivie:1983ip,Krauss:1985ub,
  Sikivie:1985yu,Asztalos:2009yp,Kahn:2016aff,
Brubaker:2016ktl,Baryakhtar:2018doz,Andriamonje:2007ew,
Anastassopoulos:2017ftl,Schlattl:1998fz,Armengaud:2014gea,
Graham:2015ouw,Rybka:2014cya,Arvanitaki:2017nhi,Budker:2013hfa}.  Our
focus in this paper is on these axion-like particles, in particular
hyperlight axiverse axions~\cite{Arvanitaki:2009fg,Demirtas:2018akl},
which we will henceforth simply refer to as axions.
In fact, in string theory
constructions of the
axiverse~\cite{Cicoli:2012sz,Demirtas:2018akl,Halverson:2019cmy,Halverson:2019kna},
there are typically many axions with masses far smaller than the Hubble
scale today.

The second ingredient, axion strings, are generically
produced in the early universe via a mechanism called the Kibble
mechanism~\cite{Kibble:1976sj,Kibble:1980mv,Hindmarsh:1994re}.  As an
example, consider an axion that arises as the Nambu-Goldstone boson
from the spontaneous breaking of a $U(1)$ ``Peccei-Quinn''
symmetry~\cite{axion3}.  If the early universe has a high temperature
$T \gtrsim f$, this symmetry is restored, and gets spontaneously
broken as the universe
cools. Due to
causality, different Hubble patches randomly choose a different
value for the axion field value $a$.  As a result, there are
regions of space where the axion makes a loop from $0$ to $2 \pi f$ by
chance and axion strings are formed, which 
tumble blindly as they make their way across the universe.
As long as the early
universe was hot enough, string formation is inevitable. 
The corresponding cosmological description for the formation of cosmic
strings for axions arising from higher-dimensional theories 
is not as well-understood and is an interesting open problem.

Surprisingly, the physics potential of the quantized signal from strings
in combination with equation~\eqref{eq: best
interaction} has been woefully understudied.\footnote{QCD axion
  strings have the
  unfortunate property
that they either disappear early on in the evolution of the universe
making them difficult to find, or they will overclose the
universe~\cite{Zeldovich:1974uw}.}  In this article, we rectify this
oversight.  The main effect of equation~\eqref{eq: best interaction} is to
cause a polarization rotation of linearly polarized light propagating
in the axion field.
In the case of axion ambient density, this effect has been 
well-studied~\cite{Harari:1992ea,Lue:1998mq,Pospelov:2008gg,Kamionkowski:2008fp,
Ivanov:2018byi,Fujita:2018zaj,Caputo:2019tms,Fedderke:2019ajk},
however, the reach is always limited by the fact that typically
$a/f\ll 1$ for dark matter axions, and that the total polarization
rotation angle depends only on the initial and final field value of the
axion ($\Delta a/f \ll 1$) along the path of photon propagation.  In
the presence of an axion string, there is a dramatic effect as $\Delta
a/f = 2 \pi$ when one goes around a string regardless of whether or
not the axion is dark matter and regardless of the size of $f$. Thus,
photons whose paths differ by a loop around an axion string have a
relative polarization rotation angle which is quantized in units of
$\anomaly\alphaem$. This
is analogous to performing an Aharanov-Bohm experiment with photon
polarization.  Additionally, such a polarization
rotation angle adds up as the number of strings grow -- contrary to the
expectation that it should only depend on the initial and final values
of the axion field. This is due to
the presence of topological defects which break the condition that
axion field is continuous (see~\cite{Fedderke:2019ajk} for a
comprehensive analysis). 

The Cosmic Microwave Background (CMB) is a valuable tool to study
axion strings.  The CMB provides a backlight which lights up the axion
strings.  The linearly polarized light of the CMB is rotated by
$\mathcal{O}(\alphaem) \sim 1\%$ by an axion string, providing a
unique opportunity by which
to look for axion strings.  This is particularly exciting given that
the CMB is currently sensitive to effects on the order of
$1\%$~\cite{Akrami:2018odb,Aghanim:2019ame,Ade:2014afa,Keisler:2015hfa,Array:2017rlf,Ade:2018iql,Pogosian:2019jbt}.
The polarization measurement of the CMB is expected to improve
dramatically in
next generation CMB
measurements~\cite{Abazajian:2016yjj,Pogosian:2019jbt,Ade:2018sbj,Matsumura:2016sri,Hanany:2019lle}.
We will show that these experiments will be able to probe {\it any}
axion string for all axion masses less than $10^{-27}$ eV.  In cases
where there are additional topologically stable objects like
domain-walls (DW), the resulting string-domain wall network can also be
probed. This CMB birefringence signal can thus probe a vast range of
the parameter space.

We 
discuss the effect of a single
axion string on photon propagation by inducing a quantized rotation of
photon polarization in section~\ref{sec:singlestringpol}.  In
section~\ref{sec:implications}, we discuss the
expectation on the size of such a polarization rotation angle from
theoretical considerations and the implications of measuring this
phase on our understanding of charge quantization.
In section~\ref{Sec: pol}, we review the current status on the evolution of string networks 
and provide both
analytical estimates and numerical results for the power spectrum of
the polarization rotation angle of CMB photons caused by a network
of strings. In section~\ref{Sec:
CMB} we discuss how to use the CMB polarization measurements to look
for axion strings.  In section~\ref{Sec: other} we discuss other means by
which axion strings can be detected. Finally, we conclude in
section~\ref{Sec: conclusion}.

\section{Photon polarization rotation by a single axion string}
\label{sec:singlestringpol}

The axion is a pseudo-Nambu-Goldstone boson (PNGB) of a spontaneously
broken global $U(1)$ symmetry.  Due to its nature as a PNGB, the axion
is a compact field with a fundamental period, $a = a+2\pi f$.  The
global $U(1)$ is not exact and can be broken by gravity and/or
anomalies. These effects produce an exponentially small potential for
the axion, which breaks the continuous $U(1)$ symmetry down to a
discrete shift symmetry. This remaining discrete shift symmetry may or
may not be the fundamental period of the axion, and hence can lead to
a left-over $Z_{N_{\rm DW}}$ symmetry, where ${N_{\rm DW}}$ is the domain wall number. In general this $Z_{N_{\rm
DW}}$ symmetry itself may not be exact, but any breaking of this
symmetry can easily be even more suppressed than the axion mass
itself.

If the reheating temperature of the universe is high enough that the
$U(1)$ symmetry is restored, then the subsequent $U(1)$-breaking phase
transition produces topological defects by the Kibble mechanism.  When
the axion mass is negligible (e.g. for $H > m_a$), there are only
string-like excitations.  These axion string networks form at the
$U(1)$ phase transition and rapidly approach a scaling solution of
only a few strings per Hubble patch due to string
interactions~\cite{Kibble:1976sj,Kibble:1980mv,Vilenkin:1981kz,Vilenkin:1982ks,Vilenkin:1984ib,Gorghetto:2018myk,Buschmann:2019icd}.
Once $H < m_a$, domain walls connecting the strings form. 
If $N_{\rm
DW}=1$, these domain walls pull the strings together and quickly
convert all of the energy density in domain walls and strings into
axion particles.  If $N_{\rm DW} \ne 1$, then the string/domain wall
network is stable.  If the $Z_{N_{\rm DW}}$ is not exact, then the
$Z_{\rm N_{DW}}$ breaking effects cause regions in the false vacuum to
shrink, and eventually the string/domain wall network is destroyed~\cite{Kamionkowski:1992mf}, which can potentially take a very long time.

In this section we outline the physical effect relevant for this
paper, the rotation of photon polarization by axion strings. For this
purpose, we focus on the effect of a single string, and we postpone
the analysis of the string network to section~\ref{Sec: pol}.
\begin{figure}[tp]
\centering
\fbox{\parbox[c][12em]{11em}{\hfill$\hbox{ \convertMPtoPDF{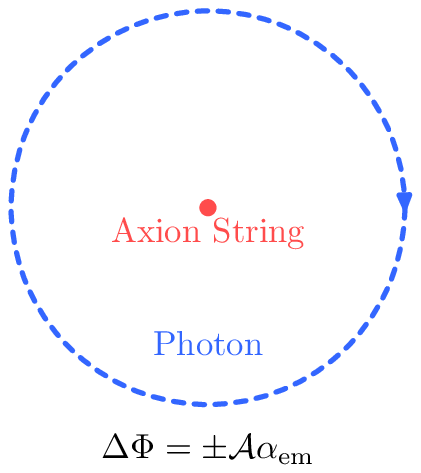}{.8}{.8} }$\hfill\,}}
\fbox{\parbox[c][12em]{16em}{\hfill $\hbox{ \convertMPtoPDF{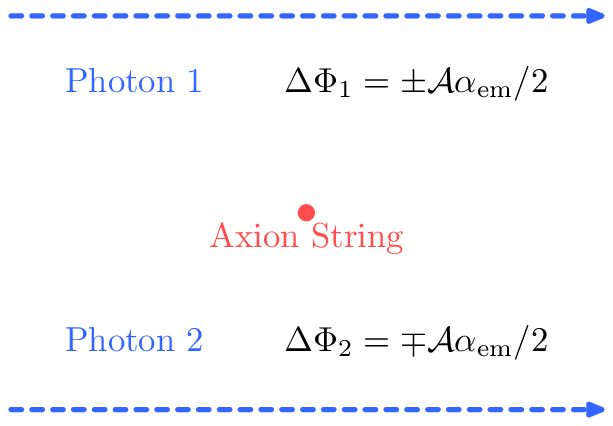}{.8}{.8} }$\hfill\,}}
\fbox{\parbox[c][12em]{14em}{$\hbox{ \convertMPtoPDF{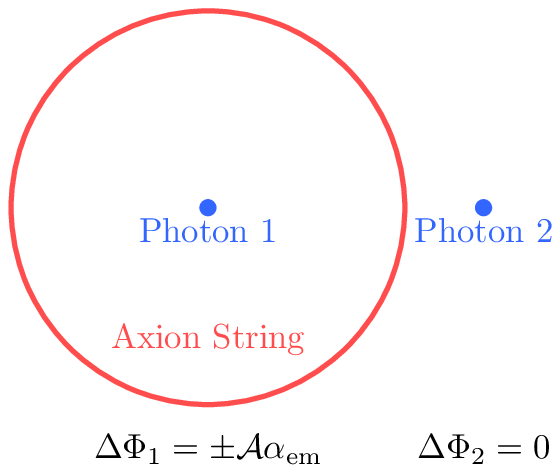}{.8}{.8} }$}}
 \caption{The observable effects of cosmic axion strings (red dots and
 red lines) on the polarization of photons passing by (blue dots or
 blue dashed lines) in various situations. The red and blue curves
 represent a string or a photon that lies on the surface while the red
 and blue dots represent a string or a photon that goes through the
 surface. (Left) Photons that go around an axion string accumulate a
 polarization rotation angle $\Delta \Phi = \pm \anomaly \alphaem$. 
 (Middle) Photons that go past an axion string from and to infinitely
 far away on the opposite side of a string accumulate a relative phase
 of polarization rotation $\Delta \Phi = \pm \anomaly\alphaem$.
 (Right) Photons that pass through an axion string loop from and to
 infinitely far away get a polarization rotation angle of
 $\Delta\Phi = \pm \anomaly\alphaem$ while photons that do not get 0
 polarization rotation angle.}\label{fig:effect}
\end{figure}
We normalize the axion-photon coupling as,
\begin{align}
  \mathcal{L}
  &\supset
  \frac{\anomaly \alphaem}{4\pi f} a F^{\mu\nu} \widetilde{F}_{\mu\nu} ,
\end{align}
where as mentioned before, $\anomaly$ is the anomaly associated with
the axion, and $\widetilde{F}_{\mu\nu} = \frac12
\epsilon_{\mu\nu\alpha\beta}F^{\alpha\beta}$. 
If the axion $a$ is a constant, then this term has no
effect in the absence of monopoles or non-trivial topology.  If the
axion is oscillating in time or
has a gradient in space, then linearly polarized light passing by
undergoes a polarization rotation by the amount (see
reference~\cite{Fedderke:2019ajk} for a careful derivation),
\begin{align}
  \Delta \Phi 
  &= 
  \frac{\anomaly \alphaem}{2\pi} \frac{\Delta a}{f} .
\end{align}

Around a string, the axion has a field profile
\bea
\frac{a}{f} = \pm \theta + \, \text{constant} ,
\eea
where $\theta$ is the angle around the string and the $\pm$ depends on the direction of the string. Therefore, as
shown in figure~\ref{fig:effect} (left), photons that go around an
axion string in a closed loop accumulate a polarization rotation of
\bea
\Delta \Phi = \pm \frac{\anomaly \alphaem}{2\pi} 2 \pi = \pm \anomaly
\alphaem.
\eea
Such a phase $\Delta \Phi$ exists because of the discontinuity of
the axion field $a$ at the core of the string (the axion field is
periodic but not continuous)\footnote{This effect is similar to the 
Aharonov-Bohm effect where the derivative of the axion
$\partial_{\theta} a$ in our case plays the role of the gauge field
$A_{\theta}$.}. This {\it
geometrical phase} is topological in nature, and therefore independent
of the details of the photon trajectory or the shape of the axion
string. 

A similar conclusion can be reached for a geometry which is more
directly related to the CMB. 
Photons which pass by the left of a string from
and to infinity see $\Delta a/f =\pm \pi$ while photons which pass by
the right of a string see $\Delta a/f = \mp \pi$. As shown in
figure~\ref{fig:effect} (middle), the two photon trajectories
accumulate a relative polarization rotation angle of\footnote{The relative polarization rotation angle accumulated by the two photon trajectories would still be $\Delta \Phi = \pm \anomaly \alphaem$ if the two photon trajectories originate from infinity, pass by the string on opposite sides, and end at the same point.}
\bea
\Delta \Phi =\Delta\Phi_1-\Delta \Phi_2 = \pm \anomaly \alphaem .
\eea

Finally, there is the case of finite string loops, as shown in
figure~\ref{fig:effect} (right). When a photon misses a
string loop, $\Delta a/f = 0$, but if a photon passes through a
string loop, $\Delta a/f = \pm 2 \pi$ where in this case the $\pm$
comes from if the loop is right or left handed.  Thus, if a photon
passes through a string loop it picks up a polarization rotation of
$\anomaly \alphaem$.

The final objects that the strings may pass through are the domain
walls.  A domain wall is characterized by $\Delta a/f = 0$ outside of
the domain wall.  Passing through a domain wall has $\Delta a/f = \pm 2
\pi/N_{\rm DW}$.  So that the polarization rotation angle of a photon
passing through a single domain wall is
\bea
\Delta \Phi_{\rm DW} = \pm \frac{\anomaly \alphaem}{ N_{\rm DW}} .
\eea
Since each axion string is attached by $N_{\rm DW}$ domain walls, the
overall polarization rotation around an axion string remains
${\anomaly \alphaem}$.

In all of these scenarios, the polarization rotation angle can be positive
or negative.  This means that a photon which passes by many strings
will undergo a random walk of its polarization angle.  Therefore we
naturally expect that any signal of the strings would be proportional
to the square-root of the number of strings $\sqrt{N_s}$. We will
study the effect of a network of strings and domain walls in
section~\ref{Sec: pol}.

\section{Quantized anomaly coefficient and its implications}\label{sec:implications}

A particularly long lasting puzzle in physics is the quantization of electric charges and, in particular, the size of the minimal quanta of charge. Since Millikan's experiment~\cite{1913PhRv....2..109M}, it has been established that all the IR degrees of freedom in the Standard Model (protons, neutrons and electrons) have electric charges under $U(1)_{\rm em}$ that are integer multiples of the charge of the electron. However, the gauge group $U(1)_{\rm em}$ comes from the breaking of $SU(2)_{\rm L}\times U(1)_{\rm Y}$, while the neutrons and protons are products of the confinement of the $SU(3)_{\rm C}$. The UV degrees of freedom, quarks and leptons ($Q,\, u^c,\, d^c,\,L,\,e^c$), can have fractional $U(1)_{\rm Y}$ charges and fractional $U(1)_{\rm em}$ charges after electroweak symmetry breaking.  The very curious and somewhat fragile charge assignment of the Standard Model matter fields under the gauge group $SU(3)_{\rm C} \times SU(2)_{\rm L}\times U(1)_{\rm Y}$ is such that all of the low energy degrees of freedom have charges that are integer multiples of the electrons, which has been taken as a rather strong indication of Grand Unification~\cite{Georgi:1974sy}. In fact, simply adding to the Standard Model a vector pair of quarks with charges $(3,1,0)$ under the $SU(3)_{\rm C}\times SU(2)_{\rm L}\times U(1)_{\rm Y}$ gauge group would already lead to mesons and baryons with $1/3$ electric charge under $U(1)_{\rm em}$.

Topological defects have long been known to offer a unique opportunity to discover the unit of charge. The most famous example is the magnetic monopole~\cite{Dirac:1931kp}. A measurement of a single magnetic monopole with charge $2 \pi/e$ would be enough to prove that electrons have a unit charge under $U(1)_{\rm em}$~\cite{Preskill:1984gd}.  
In this paper, we consider a massless axion (mass smaller than $H_0$ unless otherwise specified) that couples to the SM photon.
While a discovery of an axion string provides less information than the discovery of a monopole, we can still learn very interesting information.   When light goes around an axion string, it picks up a phase rotation
\bea
\Delta \Phi = \anomaly \alphaem,
\eea
where $\anomaly$ is the UV anomaly coefficient.  Because this value is
only sensitive to the UV, $\anomaly$ can be used only to teach us about the
charge quantization in the UV. 
Namely, we cannot make any statements about whether or not the
electron is the object with the smallest charge in the asymptotic IR
or not. Rather, we
can only discuss charge quantization in the UV where effects such as
confinement are irrelevant. However, if the axion is naturally
completely massless, then all of the dynamics of the underlying
theory, most importantly, confinement, would be effectively UV
dynamics and the coefficient $\anomaly$ would inform us about charge
quantization at effectively any energy scale. Alternatively, if the
axion gets a mass from confinement of a non-Abelian gauge group, then
the axion is sensitive only to the smallest charge before confinement
of this gauge group.

The measurement of $\anomaly$ is
important because it can 
inform us about an ambiguity of the discrete symmetries in the
standard model gauge group.
The gauge group is
$SU(3)\times SU(2)\times U(1)_Y/\Gamma$, where $\Gamma$ can be
$1$, $\mathbb{Z}_2$, $\mathbb{Z}_3$ or $\mathbb{Z}_6$.  Only certain
values
of $\anomaly$ are compatible with modding out by the various discrete
symmetries and a measurement of $\anomaly$ (e.g. $\anomaly = 1/4$ could
conclusively show that the gauge group is not modded out by these
accidental discrete symmetries).  Incidentally, measurements of this
sort would also rule out minimal Grand Unified theories.

\subsection{Model-independent implications}
We first discuss what statements about the UV can be made with absolutely no assumptions.  The axion couples as 
\bea
\frac{e^2}{16 \pi^2} \left ( \anomaly \frac{a}{f} + \theta \right ) F
\widetilde F,\label{eq:electroncharge}
\eea
where $e$ is the electric charge of the electron. Going around a string, $a \rightarrow a + 2 \pi f$.  Because one ends up at the same position after performing a loop, the theory must be symmetric under this transformation.
From this, we discover that the theory is symmetric under
\bea
\label{Eq: thetaperiodic}
\theta \rightarrow \theta + 2 \pi \anomaly .
\eea
In a simply connected non-abelian theory, the $\theta$-angle must be
$2 \pi$ periodic and so
$\anomaly$ must be an integer.\footnote{In a non-simply connected
non-abelian theory, e.g. $SU(N)/\mathbb{Z}_N$, $\theta$ is $2 \pi N$
periodic and a stronger statement can be made that $\anomaly$ must be
an integer multiple of $N$.}  For a compact abelian theory, $\theta$ can be
periodic with any rational number times  $2 \pi$.
Combined
with equation~\eqref{Eq: thetaperiodic}, this shows that $\anomaly$ is
a rational number, which has information about the value of the fundamental
charge~\cite{Callan:1984sa}. For a non-compact gauge group such as
$\mathbb{R}$, $\anomaly$ is not quantized.

A massive axion  can get IR contributions to
its coupling to photons from mixing with mesons of a different sector. 
Let us consider as a test example the QCD axion in the two-quark limit with
the heavy $\eta'$ integrated out. 
This example should only be taken as an illustration of the effect of mixings.  We are using the example of the QCD axion for reasons of familiarity, however QCD axion strings cannot be searched for using the methods described in this paper.
The same argument applies to any two pseudo-goldstones that undergo mixing. 
The QCD axion mixes with the $\pi^0$ in chiral perturbation theory through the interaction
\bea
V \propto -m_u \cos \left( \frac{\pi^0}{f_\pi} - \frac{a}{2 f_a} \right) - m_d \cos \left( \frac{\pi^0}{f_\pi} - \frac{a}{2 f_a} \right) \, ,
\eea
where $m_{u,d}$ are the light quark masses and $f_\pi$ is the pion
decay constant. 
For any given value of the QCD axion, the pion relaxes to an expectation value given by 
\begin{align}
  \left\langle\frac{\pi^0}{f_\pi}\right\rangle
  &=
  \tan^{-1}
  \left(
    \frac{m_u-m_d}{m_u+m_d}
    \tan \left(\frac{a}{2f} \right)
  \right)
  \label{eq:pionaxion}
  \, .
\end{align}
From this, one can see that as the QCD axion field value changes from $0$ to $2 \pi f_a$ that the $\pi^0$ field also develops a non-trivial field
profile and itself also changes from $0$ to $-\pi f_{\pi}$ around
the string.
The QCD axion and the pion have anomalies
\begin{align}
  \mathcal{L}
  &=
  \frac{\anomaly_a \alphaem}{4\pi} \frac{a}{f_a} F \widetilde F + \frac{\anomaly_\pi \alphaem}{4\pi} \frac{\pi^0}{f_\pi} F \widetilde F .
  \label{eq: best interaction2}
\end{align}
As a photon goes around a string, it experiences a $\anomaly_a$ rotation due to the QCD axion's anomaly and a $\anomaly_\pi/2$ rotation due to the pion anomaly.  Because both of these are rational numbers in the charge basis, the sum is also a rational number.
This argument ensures that the phase $\Delta \Phi$ around a string is the same at all distances from the string core. 
Of course, one normally works in the mass basis where the mixing
between the $\pi^0$ and QCD axion is an irrational number.  Thus, the QCD axion
and pion both acquire an irrational anomaly in this basis, but the net
result is still a rational anomaly coefficient as the end result does not depend on which basis one chooses.
The polarization rotation angle $\Delta
\Phi$ around a QCD axion string gets affected by the mixing between
axion and pions only to the extent that the net result is the sum of
the contributions of the two anomalies $\Delta \Phi = (
\anomaly_a + \anomaly_\pi/2) \alphaem$.  Similarly, the polarization rotation angle
$\Delta \Phi$ of a string of a generic axiverse axion does not depend
on the mass and mixings of the axion. The relation in equation
\eqref{eq:pionaxion} is, in general, independent of which of the two
pseudo-Goldstones is heavier. Therefore, the same conclusions apply to
mixings with lighter particles.

The periodicity of $\theta$ can be used to constrain the value of the
fundamental charge. This can be seen by the Witten
effect~\cite{Witten:1979ey} as a monopole rotates around an axion
string (see~\cite{Fischler:1983sc, Sikivie:1984yz, 
  Davis:1986qwa, Wilczek:1987mv} for a description of how charge
conservation works in this system).  A particle with magnetic charge
$g$ has an
electric charge
\bea
q_e = \left ( \frac{e g}{2 \pi} \right) \frac{\theta e}{2 \pi}
\eea
Quantization of electric charge says that
\bea
g = \frac{2 \pi}{e_{\rm min}} .
\eea
Note that if there existed only a single type of monopole, then the
spectrum is not invariant as $\theta$ changes over its period.  On the
other hand, if
there are an infinite number of monopoles with electric charges
uniformly spaced, then as $\theta \rightarrow \theta + \rm{period}$, the
monopoles can be exchanged and the spectrum is invariant.  Thus, the periodicity of $\theta$ gives us the difference in electric charges of the monopole spectrum.

To see how the periodicity can give us information about the
fundamental charge $e_{\rm min}$, say we discover that $\theta$ is
periodic with period  $2 \pi/n^2$, where $n$ is an 
integer.\footnote{Our normalization of
  $\theta$ is with respect to the electron charge, as defined in
equation~\eqref{eq:electroncharge}. }  From
this, we can see that the monopole spacing is
\bea
\Delta q_e = \frac{e}{e_{\rm min}} \frac{e}{n^2}
\eea
Assuming conservatively that the spacing between monopoles is the
minimal electric charge (namely we assume that $\theta$ is $2 \pi/n^2$
periodic rather than having an even smaller period), we find that
\bea
e_{\rm min} = \frac{e}{n}
\eea
so that the electron is not the minimally charged object in this theory.

\subsection{Model-dependent implications}

When obtaining an anomaly from integrating out new fermions in the UV, one obtains
\begin{align}
\anomaly  = \sum_f Q_{a,f} Q_f^2
\end{align}
where $Q_{a,f}$ and $Q_f$ are the ``PQ'' and electric charges of the fermions.
The ``PQ'' charge of a fermion is necessarily an integer for a string
that is properly normalized.  If $\anomaly$ is measured to be a
fraction, then there must exist other particles that have fractional charge.
Since the SM itself has particles with fractional charge $1/3$ and
$2/3$, a natural expectation is that if the UV theory is SM-like, then
$\anomaly$ will be a multiple of $1/9$.
This expectation is realized in theories such as minimal Grand Unified
Theories ($\anomaly = 4/3$). If the axion is completely massless
naturally, we expect this coefficient $\anomaly$ to be an integer
multiple of the smallest charge of gauge-invariant asymptotic 
states.\footnote{Adding a vector pair of quarks with charge $(3,1,0)$ under
the $SU(3)_{\rm C}\times SU(2)_{\rm L}\times U(1)_{\rm Y}$ gauge group
would lead to mesons and baryons with $1/3$ electric charge under
$U(1)_{\rm em}$. In this theory, it is possible to arrange for the
axion to be massless while the coefficient $\anomaly$ to not be an
integer.} 
In the parameter space we can probe, the axion is nearly massless.
Therefore, if it couples to a non-abelian gauge group, this gauge
group must confine at extremely low scales
(see~\cite{Hook:2018jle} for the only known counter-example). This
would require
a delicate cosmological history for the dark sector.

The minimal polarization rotation angle due to a single string with a
SM-like UV theory is $\alpha_{\rm em}/9$. This is a reasonable
sensitivity goal for current and future CMB experiments, and with
CMB-S4~\cite{Abazajian:2016yjj} the anomaly coefficient can be
measured to good precision.  If $\anomaly$ is measured to be some
other rational number, then it would suggest that the quantum of
hypercharge is smaller than expected in the UV theory.  If $\anomaly$
is irrational---which would be hard to prove experimentally but
suggestive if it is not a rational number with small integers---it
would point to either that the underlying gauge group of
electromagnetism is non-compact (although there are strong arguments
against this possibility~\cite{Banks:2010zn}), or the existence of
kinetic mixing of the photon with a dark photon.

\subsection{Kinetic mixing with a dark photon}

Kinetic mixing between the photon and a dark photon ($A'_{\mu}$) can be
generated by integrating out UV fermions, resulting famously in a
small irrational millicharge at loop order without the need to invoke
explicit violation of charge quantization. Depending on whether the
axion is coupled as $a F \widetilde{F}'$ (mixed anomaly) or as $a F'
\widetilde{F}'$ ($U(1)'$ anomaly) to the photon and dark photon, the
polarization shift around an axion string can be 1-loop or 2-loop
suppressed compared to the case of an $a F \widetilde{F}$ coupling. 
In either case, an irrational $\anomaly$ is generated.
Both cases can be looked for with future CMB experiments. Unlike the normal
case where a dark photon decouples in the massless limit
(see~\cite{Graham:2014sha} and references within), coupling to the
axion strings eliminates the freedom to rotate the dark photon field
($A'_{\mu}$) to remove the kinetic mixing.\footnote{In the case of a
$U(1)'$ anomaly, it is easy to see the effect in the basis where
only $A'_{\mu}$ couples to the axion string, kinetic mixing is
removed, and the electron is millicharged under $U(1)'$.} As a result,
the axion strings also offer a unique possibility to search for
massless dark photon in the string
photiverse~\cite{Arvanitaki:2009fg}.

In summary, we expect that the polarization rotation angle $\Delta
\Phi$ around a string to be most likely a rational number $\anomaly$ multiplying
the fine structure constant $\alpha_{\rm em}$, which is independent of
mass and kinetic mixing of the axiverse axions. This rational number
$\anomaly$ can contain rich information about charge quantization and
the structure of the UV theory.  Even more information can be gleaned
if we observe different values of $\anomaly$ coming from different types of
axion strings corresponding to different axions.

\section{Polarization rotation by a string network} 
\label{Sec: pol}
In this section, we study the effects of a string network on the
polarization of CMB photons.  We begin by reviewing
properties of string networks derived from numerical studies. We then
proceed to show the theoretical estimates and simulation results for
the polarization rotation angle power spectrum caused by such a network in
the universe between the time of CMB and today.   At the end of the
section, we will comment on the validity of our simple analysis and
point to directions where detailed simulations of string network
evolution can be helpful.  We treat the axion as completely massless
(lighter than $H_0$ practically), though a massive axion with domain
wall number $N_{\rm DW} \neq 1$ does not qualitatively change the
prediction.

\subsection{Properties of the string network}
\label{Sec: string basics}

The string network for axions has been studied extensively in the
literature~\cite{Gorghetto:2018myk,Buschmann:2019icd}.  There are also studies of domain wall/string network,
but since this easily leads to an overabundance of
dark matter when axions make up the dark matter, this scenario is less
well-studied~\cite{Saikawa:2017hiv}.  The scenario we are concerned
with involves the so-called global strings, where the string tension is
\bea
\mu \approx \pi f^2 \log \left( \frac{m_r}{H} \right)
\,.
\eea
Here $m_r$ is the mass of the radial mode ($1/m_r$ is the size of the
string core).  In what follows, we highlight some of the
qualitative features that recent numerical simulations of global strings have found~\cite{Gorghetto:2018myk,Buschmann:2019icd}.

\paragraph{\underline{Scaling limit and scaling violation}}
Regardless of how the strings were produced, a string network quickly
evolves towards an attractor solution called the scaling
limit~\cite{Kibble:1976sj,Kibble:1980mv,Hindmarsh:1994re}.  
The energy density in the string network is parametrized as
\bea
\label{Eq: energy}
\rho_{\rm strings} = \xi(\eta) \mu H^2,
\eea
where $\eta$ is the conformal time, and $\xi(\eta)$ measures the  length of strings per Hubble patch in Hubble units.\footnote{Note a small numerical difference in terms of definition of $\xi$ as compared to~\cite{Gorghetto:2018myk}.}. The scaling limit has $\xi(\eta)$ equal to a constant independent of time.
Numerically, it has been found that there is a logarithmic deviation
from the scaling solution~\cite{Gorghetto:2018myk}
\bea
\xi(\eta) \simeq \alpha \log \left( \frac{m_r}{H} \right)
+ \beta \qquad \alpha, \beta \sim 0.1\mathrm{-}1 .
\eea
The range comes from different simulations giving different results.
For our studies, we take $\xi$ to be a constant after recombination,
and in the broad range,
\bea
10^3 \gtrsim \xi \gtrsim 1 
\,,
\eea
neglecting the very mild logarithmic variation between the CMB epoch and
today.

Importantly, the energy density in the string network diluting away as 
$\xi \mu H^2$ has the consequence that their energy dilutes away like the
dominant
form of energy in the universe.  They decay away as radiation (matter)
during radiation (matter) domination, so that they
do not come to dominate the total energy density and can be
present throughout much of the history of the universe without
dramatically affecting cosmology.

\paragraph{\underline{String Length}} From equation~\eqref{Eq: energy}, one can
see that the total string length in a Hubble volume is roughly $L_{\rm string} \sim
\xi/H$.  It has been found that roughly 80\% of the string length is
contained in strings longer than Hubble, while only 20\% of the string
length is contained in strings with length smaller than Hubble.  As
expected from a scaling solution, the smaller strings follow a
scale invariant distribution with $L_{\rm string} \, {\rm d}n_L/{\rm d} L_{\rm string}$ being a constant.
Because small strings are only a small part of the total energy
density, it is useful to treat the situation as having roughly
$\xi(\eta)$ Hubble length strings per Hubble volume.  Despite calling
them Hubble length strings, these strings are connected and form
infinitely long strings. One other important length scale is the  radius of curvature of the string. Due to causality, the radius of curvature of these strings has to be roughly the Hubble radius.
Therefore, we can
treat these strings as either infinitely long strings or as Hubble sized loops of strings.

\paragraph{\underline{Domain Walls}} Domain walls form when $H \sim m_a$ and have
a tension $\sigma \sim m_a f^2$.  The domain walls that form end on
strings and the response of the scaling solution of strings to domain walls
depends critically on $N_{\rm DW}$.  If $N_{\rm DW} = 1$, then there
is a single domain wall that ends on a string and it pulls the strings
together and causes them to disappear within a few e-folds.  All of
the ambient energy in the domain walls and strings is transferred into
the axion particles.  Because this behavior destroys the signal
associated with strings, we take either $m_a$ small or $N_{\rm DW} >
1$.

If $N_{\rm DW} \ne 1$, then attached to each string there are $N_{\rm
DW}$ domain walls.  When crossing each domain wall, $a/f$ changes by
an amount $2 \pi/N_{\rm DW}$.  Because the domain walls are all
pulling in different directions, the string-domain wall network is
stable.  As before, there remains $\sim \xi$ strings per Hubble
volume~\cite{Hiramatsu:2012sc,Hiramatsu:2013qaa}.  However, unlike
before, most of the energy of the system is now in domain walls ($\xi
N_{\rm DW} m_a H f^2$ versus $\xi \mu H^2$).  Note that because the
majority of energy density is no longer stored in the strings but is instead stored in
the domain walls, the total energy density in the system dilutes away
very slowly and can overclose the universe if $m_a$ or $f$ is not
small enough.

\paragraph{\underline{Ambient Axion density} }
A string by itself dilutes away
slower than radiation or matter.  However as mentioned before, the
string network dilutes away as fast as radiation or matter depending
on the ambient background.  The extra energy is radiated into
relativistic axions with energy $\sim 10 H$~\cite{Gorghetto:2018myk}.
This ambient energy density results in a field value $a(x,t)$ that is
oscillating in time.  The energy density in axions scales as $\pi \xi
f^2 H^2 \log \left ( m_r/H \right )$, where $m_r \sim f$ is the mass
of the radial mode.  This energy density in axions results in a field
value 
\begin{align}
  \left\langle \frac{a^2}{f^2} \right\rangle 
  &\sim 
  \frac{1}{100} 
  \xi(\eta) \log \left (\frac{m_r}{H} \right)
  \,.
\end{align}
The exact numerical coefficient will
depend significantly on the spectrum of the emitted axions and is a
subject of active research~\cite{Gorghetto:2018myk,Buschmann:2019icd}.
The effect of an ambient axion density on the polarization rotation has
been studied
in~\cite{Lue:1998mq,Pospelov:2008gg,Ivanov:2018byi,
Fujita:2018zaj,Caputo:2019tms,Fedderke:2019ajk},
which can cause a smearing of the polarization at the time of
recombination.  The ambient axion density radiated by the string
network can also cause a spatially varying unquantized
polarization rotation angle, which can affect our CMB measurement. We
will comment on the effect of this ambient axion density on our signal
at the end of section~\ref{Sec: pol}.

\subsection{Analytical approximation of the two-point function}

There are two basic polarization rotation effects present in the CMB coming from strings.
The first effect is the discontinuity that occurs when one compares the polarization rotation of light going to the left or to the right of the string.  The second effect is the continuous change in polarization that comes from going part way around the string.  Rotations arising from strings at high redshifts are dominantly discontinuous as the CMB has passed by the string entirely, so that $\theta_i - \theta_f = \pi$.
Rotation coming from nearby strings have a large continuous part as light has not passed the string entirely.

Under some very simple conditions, we can derive a semi-analytical
estimate for the two-point function of the polarization rotation angle. We
assume that at any given time that the string network consists of
circular loops of physical radius equal to the Hubble radius $1/H$ at
that time. The number of loops at any given time is set by the scaling
limit described in section~\ref{Sec: string basics}. Since we neglect
smaller loops, this approximation will not be accurate for angular
sizes smaller than Hubble when the CMB was produced ($\ell \sim 100$
modes).

\begin{figure}[t]
  \centering
  \includegraphics[width=0.30\textwidth]{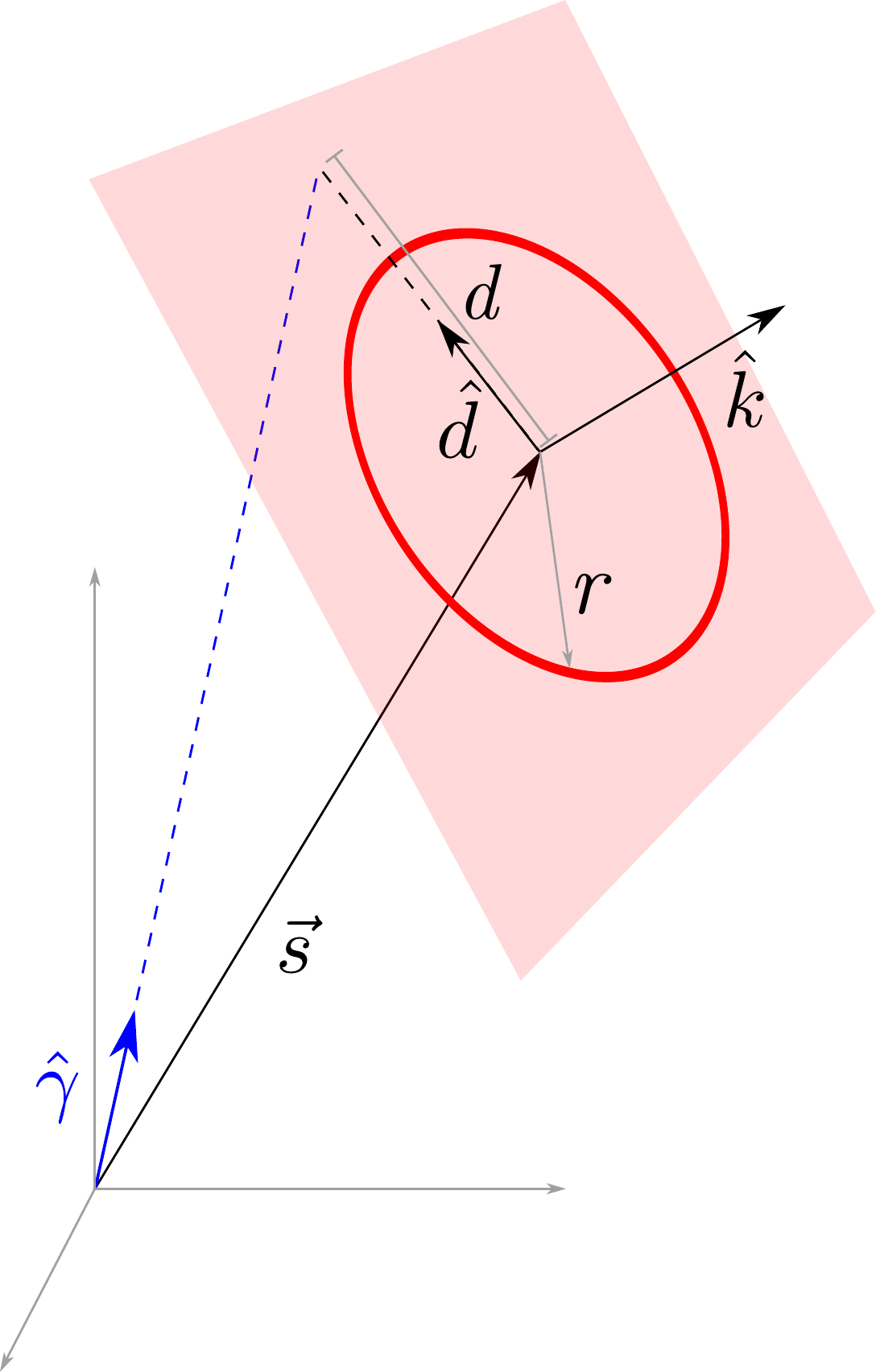}
  \caption{A schematic figure showing the the string loop in the sky
    and the various parameters used to describe it. The red circle is
    the axion string loop while the blue line shows the photon
    direction. $\vec{s}$ gives the location of the center of the
    string loop, $r$ is its radius, $\hat{k}$ is the orientation,
    $\hat{\gamma}$ is the direction the CMB photon is coming from, and
    $d$ is the distance between the photon trajectory and the loop
  center measured in the plane of the loop (red shaded surface).}
  \label{fig:loop-geom}
\end{figure}

We work in conformal time,
\begin{align}
  ds^2 = a^2(\eta)(d\eta^2 -d \vec{x}^2)
\end{align}
and describe the computation in a matter dominated universe for simplicity of analytical
expressions (the case with an accurate evolution history of the universe can be obtained by simply using the correct Hubble evolution $H(a)$).
In this case, the comoving Hubble is
\begin{align}
  a H
  &=
  \frac{2}{\eta}
\end{align}
We choose a comoving coordinate system with the observers at the
origin.
The probability density of finding a
string loop with the center of the loop at coordinates ($\vec{s}$) with orientation
($\hat{k}$)
and a coordinate radius of the string loop
($r$) at a time ($\eta$) is denoted as
\begin{align}
  P(\vec{s}, r, \eta, \hat{k})
  &=
  \delta \left(r-\frac{1}{a H}\right)
  f(\eta)
\end{align}
where we have assumed that this probability is spatially uniform (independent of the direction of $\hat{s}$) and
isotropic in $\hat{k}$. The
function $f(\eta)$ can be fixed by demanding that the total energy in
a given volume
follow the expected scaling limit:
\begin{align}
  \int dr\,d^3 \vec{s}\,d^2 \hat{k}\ 
  (2\pi r a) \mu(\eta) 
  P(\vec{s}, r, \eta, \hat{k})
  = V a^3 \xi(\eta) \mu(\eta) H^2
\end{align}
where $\mu(\eta) \approx \pi f^2 \log(f/H)$ is the string tension at
time $\eta$. This implies,
\begin{align}
  f(\eta)
  &=
  \frac{1}{4\pi}
  \frac{\xi(\eta) (aH)^3}{2\pi}
\end{align}

We are interested in the polarization rotation of a photon from a given
direction $\Phi(\theta,\phi)$. We approximate the polarization
rotation by assuming that the photon undergoes a polarization
rotation of $\anomaly \alphaem$  when passing through a loop and 0
otherwise. This is a good approximation for string loops that subtend a small angle on the sky and both the time of emission (CMB) and time of observation (today) can be treated as infinitely far away from the string loop, but we do not expect to recover the low-$\ell$ signal with this approximation.

To proceed, we need the condition for a photon from the direction $\hat{\gamma}$
to pass through a loop centered at $\vec{s}$, with a normal vector $\hat{k}$
and radius $r$. The distance of the photon trajectory from the center
of the string loop (calculated in the plane of the string loop) is 
\begin{align}
  d = \left|\frac{\vec{s}\cdot(\hat{\gamma} (\hat{d}\cdot \hat{\gamma}) -
  \hat{d})}{1-(\hat{d}\cdot{\hat{\gamma}})^2}\right| ,
\end{align}
where $\hat{d}$ is the unit vector along the radial direction of the
string in the plane containing $\vec{s}$ and $\hat{\gamma}$,
\begin{align}
  \hat{d}
  &=
  \frac{\hat{k}\times(\hat{\gamma}\times \vec{s})}
  {|\hat{k}\times(\hat{\gamma}\times \vec{s})|}
\end{align}
For the photon to pass through the string loop, we need $d < r$. 
The geometry is shown explicitly in figure~\ref{fig:loop-geom}.
The total polarization rotation is given as a line-of-sight
integral over the probability of the photon to pass through a given
loop. We can calculate the two-point function for the polarization
rotation angle as an integral over the ensemble. 
At a time $\eta$, the photon is at a distance $(\eta_0 - \eta)$ from us  and the strings that can contribute to the
two-point function at this epoch have a distance $\vec{s} = (\eta_0 -
\eta)\hat{s}$. Thus, the two-point function can now be written
analytically,
\begin{align}
  \langle
  \Phi(\hat{\gamma})
  \Phi(\hat{\gamma'})
  \rangle
  &= (\anomaly \alphaem)^2
  \int d\eta
  \int d^2 \hat{s}
  \int d^2 \hat{k} \,
  (\eta_0-\eta)^2
  f(\eta)
  \nonumber\\&\qquad\qquad\qquad\qquad\times
  \Theta\left(\frac{\eta}{2} - d(\hat{s},\hat{\gamma},\hat{k},\eta)\right)
  \Theta\left(\frac{\eta}{2} - d(\hat{s},\hat{\gamma'} ,\hat{k},\eta)\right)
\end{align}
where $\Theta$ is the Heaviside-theta function ensuring that the
two-point function only gets contributions when both photons pass
through
the same loop. The picture where we think of photons as passing through small
loops breaks down when the radius $r$ of the loop becomes comparable
to the distance $s$ to the loop. This happens at late times when
$\eta \sim \eta_0$. So we only expect this estimate to work for
earlier times, corresponding to power at larger values of $\ell$ ($\ell \gtrsim {\rm few}$). Note
that by isotropy the two-point function only depends on the angle
between $\gamma$ and $\gamma'$, or $\cos\theta = \gamma\cdot\gamma'$.
So without loss of generality we can set $\hat{\gamma'} = \hat{z}$,
and the azimuthal angle in $\hat{\gamma}$ to zero.

\begin{figure}[tp]
  \centering 
  \includegraphics[width=0.85\textwidth]{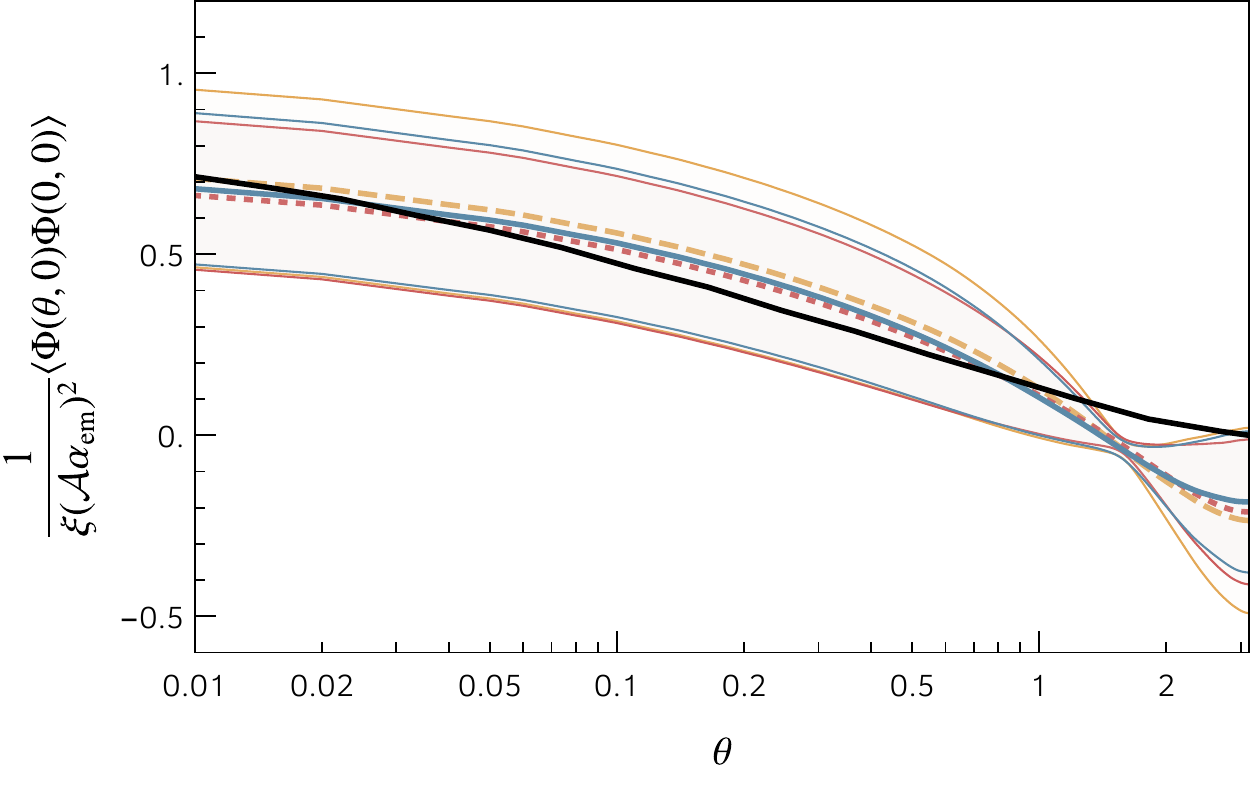}
  \caption{The two-point function of polarization rotation angle (normalized
    by $\xi \anomaly^2 \alphaem^2$). The black solid line is the analytical
    estimate calculated in the text with loops with a random normal
    direction compared to that of the photon trajectory. The blue
    solid, yellow dashed and the red dotted lines and the
    corresponding shaded regions show the central value and
    standard deviation of the two point functions from the toy simulation
  for $\xi = 1,\,\,10,\,\,100$, respectively. The uncertainties are
mainly due to cosmic variance.}
  \label{fig:twopointcompare}
\end{figure}

We present results for the two-point function computation in
figure~\ref{fig:twopointcompare} and discussion of the figure in
section~\ref{sec:limitations}.  It
is illuminating to make a further simplification to get a simple analytical
estimate for the variance of the polarization rotation angle. We assume that the
loops are all oriented perpendicular to our line of sight (i.e.~lie on
the constant $\eta$ surfaces). In this case, $\hat{k}.\hat{s} = \pm
1$. The condition for the photon to pass
through simplifies to
\begin{align}
  (\eta_0 - \eta) \tan \omega  < \eta/2.
\end{align}
where $\hat{s}\cdot\hat{\gamma} = \cos\omega$.

The polarization rotation angle along any fixed line of sight follows
a distribution where the mean is 0, and the variance is given by the
two point function with $\hat{\gamma} = \hat{\gamma'}$, or $\theta =
0$.
Parametrizing the center of the string with angles $(\theta_0,
\phi_0)$, we obtain $\omega = \theta_0$ and the two-point function does
not depend on $\phi_0$, so that
\begin{align}
\label{eq: self 2 point}
  \langle \Phi(0,0) \Phi(0,0) \rangle
  &=
  2\pi (\anomaly \alphaem)^2
  \int dc_{\theta0}  d\eta
  (\eta_0 - \eta)^2 \frac{(aH(\eta))^3 \xi(\eta)}{2\pi}
  \Theta \left(\frac{\eta}{2} - (\eta_0 - \eta)\tan\theta_0\right)
  \\&\approx \xi(\anomaly \alphaem)^2 
  \log\left(\frac{\eta_0}{\eta_{\rm CMB}}\right)
  \,.
\end{align}
In the second line we have only kept the leading $\log$ behavior.
This estimate
agrees with the intuition that a photon goes through about $\xi$ string loops 
per Hubble time in its trajectory when there are $\xi$ loops
per Hubble patch.  The polarization itself is undergoing a random walk
with $N = \xi \log\left( \eta_0/\eta_{\rm CMB} \right)$ steps and a step
size of $\anomaly \alphaem$.  This analytic estimate shows good agreement with the
numerical results.

The analytical method we described can also be used to capture the
leading behavior of the higher point correlation functions. The
higher point functions come about when multiple photons pass through
the same loop.  As a result, we expect the $N$-point correlation
functions of $\Delta\Phi$ to scale as $\xi (\anomaly \alphaem)^N$. All the odd
correlation functions, for example the 3-point function, are zero at
leading order after integrating over both orientations of the string
loops. The absence of a 3-point function and the scaling of higher
point functions with $\alphaem$ allows us to distinguish our signal and
potentially measure $\anomaly \alphaem$ and $\xi$ separately from the correlation
functions.

The above analytical analysis can be extended to the case when the axion is massive. 
When a photon passes through a domain wall $\Delta \Phi = \pm \anomaly
\alphaem/N_{\rm DW}$.  Since each string has $ N_{\rm DW} $ domain walls
attached to it, traveling around a string with or without domain
walls, one always obtains the same polarization rotation angle of $\Delta
\Phi = \pm \anomaly \alphaem$ and as a result, the previous results do not
change significantly.
When the domain wall number $N_{\rm DW}\neq 1$, the $Z_{N_{\rm
DW}}$ symmetry is exact and hence the string-domain wall network is
stable, the previous analysis still applies. In fact, as we will
discuss in more detail in subsection~\ref{sec:limitations}, the
quantized nature of the signal at CMB will actually improve if domain
walls form. The case where the $Z_{N_{\rm DW}}$ symmetry is
significantly broken or $N_{\rm DW}= 1$ is more involved. 

If the
axion mass is in the range
\begin{align}
H_0 \lesssim m_a \lesssim H_{\rm CMB},
\end{align}
the axion domain walls will form during the period when the CMB
photons are propagating to us.  The formation of the domain wall and
the subsequent disappearance of the axion string network as the
universe expands means that the larger string loops at late times do
not exist, leading to a suppression of power at large angular scales.
Practically speaking, this effect comes from limiting the $\eta$
integral to only extend to the time when the strings disappear.  The
string-domain wall network will emit most of its energy into ambient
axion density.
The exact form of the correlation functions that results from the axion radiation will depend on the spectrum of the axion radiation.  
However, we expect that the small angle correlations ($100 \gtrsim \ell \gtrsim {\rm few}$
modes of the CMB) will follow our analytic predictions.  Meanwhile,
the power in small $\ell$ modes ($\ell$ corresponding to the time when domain walls annihilate) may be affected by the axion
radiation coming from the annihilation of the string-domain wall
network.  A dedicated simulation of the evolution of the axion
string-domain wall network in this case will be extremely helpful for
predicting the signal shape.

\subsection{A toy numerical simulation}

\begin{figure}[tp]
  \centering
  \includegraphics[width=0.62\textwidth]{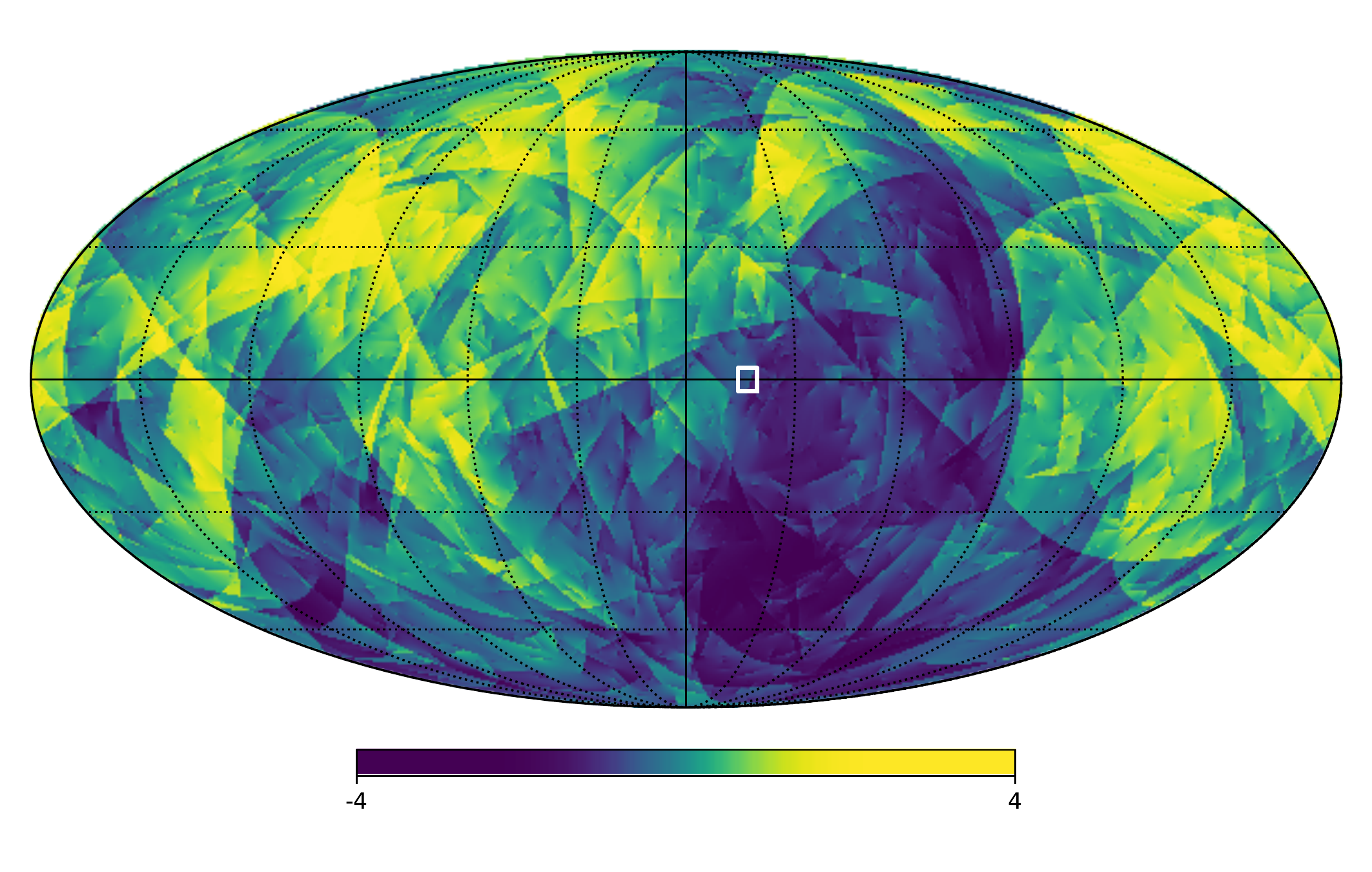}
  \\
  \includegraphics[width=0.62\textwidth]{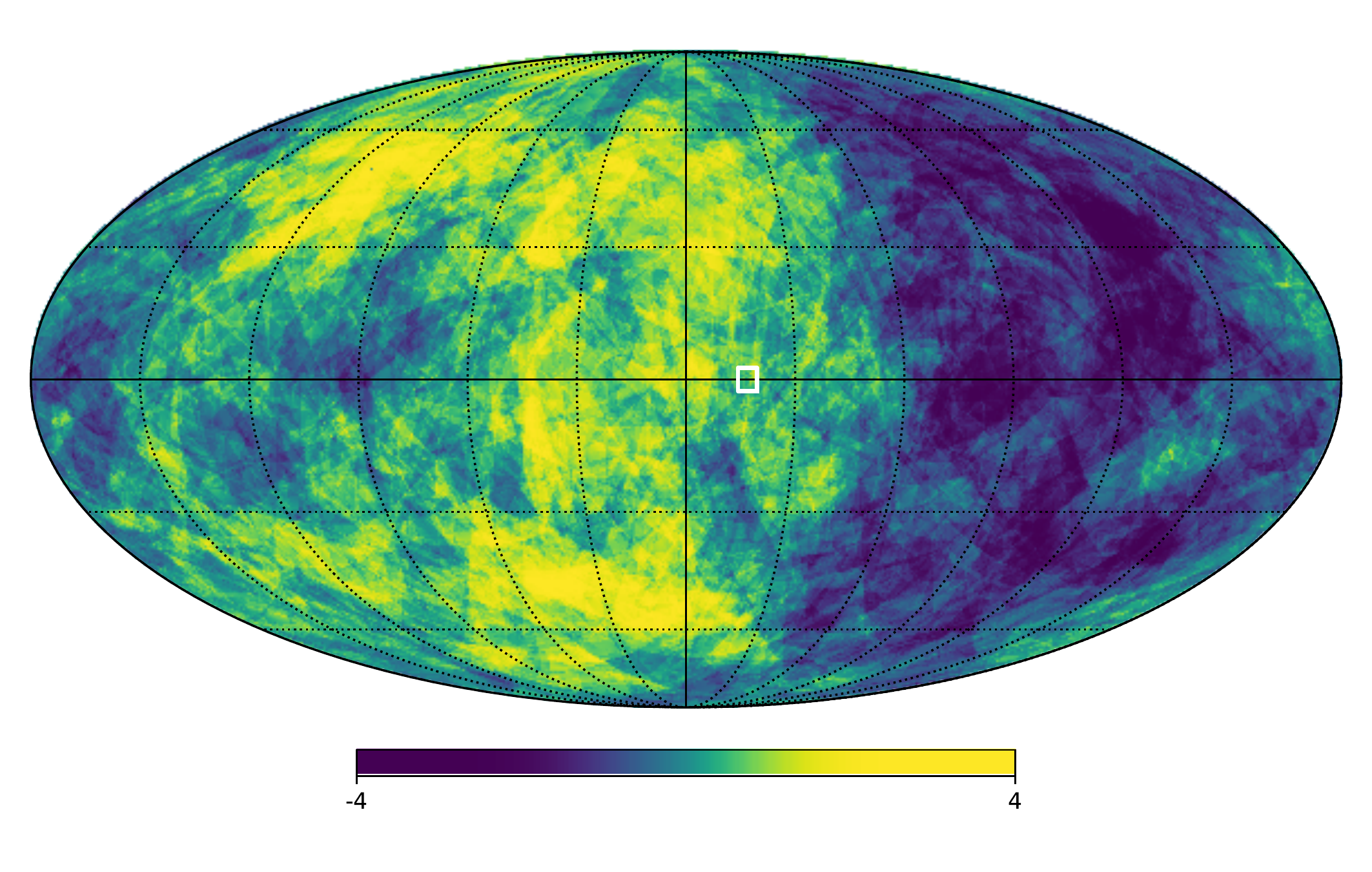}
  \\
  \includegraphics[width=0.62\textwidth]{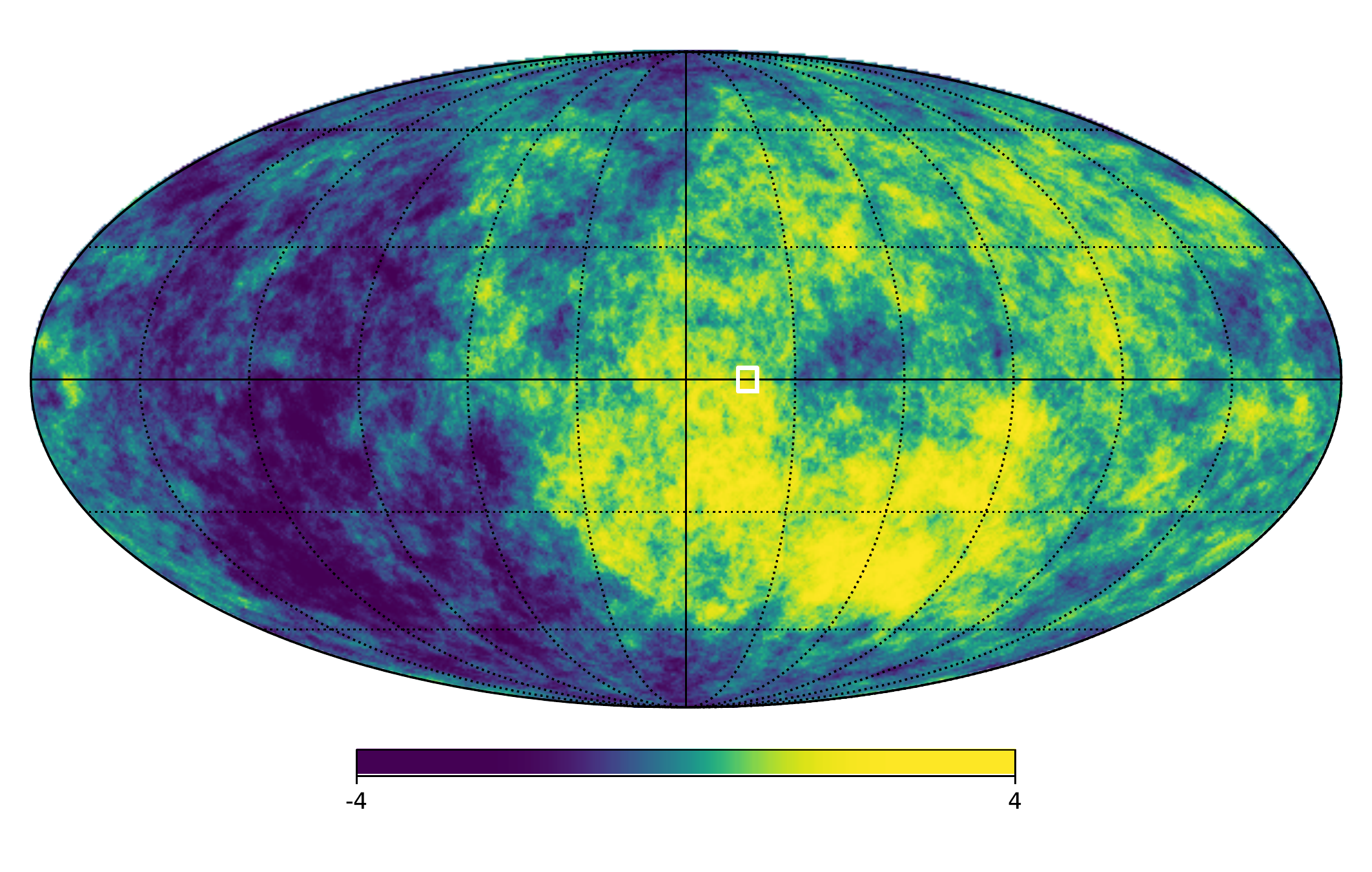}
  \caption{The polarization rotation angle map $\Phi(\hat{\gamma})$ for
  photons from a simulation of axion strings described in the text.
  The three figures, from top
  to bottom, have the number of string per Hubble patch $\xi = 1, 10, 100$ respectively. The rotation is
  normalized by $1/(\sqrt{\xi} \anomaly\alphaem)$ and is measured in radians.
The contribution from long strings today is clearly visible. The white
boxes mark the boundary of the region shown in figure~\ref{fig:disc}.}
  \label{fig:skymap}
\end{figure}

The calculation above captures the contributions from Hubble sized
loops between $z$ of a few and the CMB, but 
fails to capture the late-time behavior. In particular, there should
be low-$\ell$ signals from strings in the current universe, including
spatial features in the CMB that we can look for.
We employ a simplified
numerical simulation of strings to understand this
effect. This simulation is a toy set up that we hope captures
the qualitative properties of the signal. An obvious next step
for future work would be to use dedicated axion string network
simulations to verify the claims made in this paper. An encouraging
sign is that the results of the analytical calculation above agree
with the toy simulation for $\ell\gtrsim {\rm few}$.

The simulation assumes that all strings are straight and infinite in
length. This is an oversimplification of the string structure, which
is expected to have structure at the Hubble scale. However, this
effect is mitigated by the presence of multiple other strings with
random orientation within a Hubble volume.  Further, strings are
assumed to be static so that we do not model the string motion.
We populate a box of size $5\times(1/H_0)$ with strings at random
positions and random orientations uniformly.
In order to keep the
scaling limit as time progresses,
\begin{align}
  \rho_{\rm strings}
  &=
  \xi \mu H^2 ,
\end{align}
at every time step an appropriate number of strings are discarded at random. 

In this network of strings, we simulate photon trajectories starting
from the surface of last scattering (which we take to be a thin sphere
centered at our position) from various
angles, and calculate the polarization rotation angle along its line of
sight. This is done by assuming an axion profile around the
string to be $\nabla a = \frac{1}{r} \hat{\phi}$, where $\hat{\phi}$ is
the azimuthal direction when the string points towards the positive
$z$ direction.

\begin{figure}[tp]
  \centering
  \includegraphics[width=0.85\textwidth]{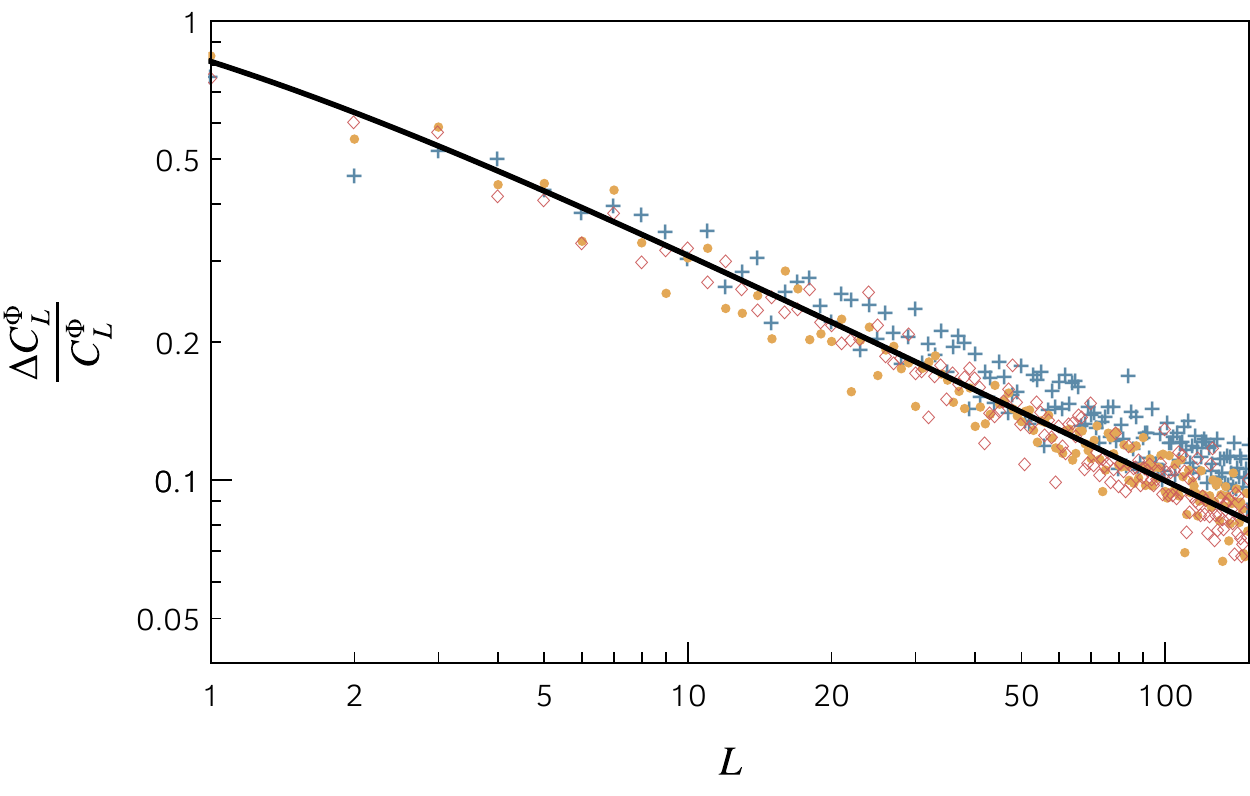}
  \caption{$\Delta C_{L}^\Phi/C_{L}^\Phi$ as a function of $L$. $\Delta
    C_{L}^\Phi/C_{L}^\Phi$ is the ratio between the standard deviation and
    the mean of $C_{L}^\Phi$ from the simulations with different choices
    of $\xi$. The blue, yellow and red markers show the value of
    $\Delta C_{L}^\Phi/C_{L}^\Phi$ from our simulations with $\xi = 1,\,\,
    10,\,\, 100$, respectively. The black solid line is the expected
    cosmic variance value of 
    $\Delta C_{L}^\Phi/C_{L}^\Phi = \sqrt{\frac{2}{2 L +1}}$. 
  These results come from 60 simulations per value of $\xi$.  }
  \label{fig:variance}
\end{figure}

We show the polarization rotation obtained in this manner in
figure~\ref{fig:skymap}. The presence of string-like structures is
clearly visible. We can further compute the two-point function and the
$C_{\ell}$ from this map, where 
\begin{equation}
C_{\ell}^{\Phi} = 2\pi\int  d \cos\theta \, \langle\Phi(\cos\theta) \Phi(0) \rangle P_{\ell} (\cos \theta)
\end{equation}
is the usual definition.
In figure~\ref{fig:twopointcompare}, we show the result and
uncertainty of the two point correlation of polarization rotation angle as a
function of angular scale $\theta$ and $\phi$ from our toy numerical
simulation. Note that the feature that all the curves approximately pass through 0
for $\theta = \pi/2$ is a consequence of the power spectrum being
dominated by the dipole in the straight string case. As our analytical
estimate shows, this suppression is not universal when the strings are
not straight.

The toy simulation also offers an opportunity to understand the
variance of the expected signal.  The overall amplitude of the signal,
as shown in figure~\ref{fig:twopointcompare}, varies from simulation
to simulation. These variations are mainly due to the effect of
strings at low-$z$, which in $\ell$-space, become variations of
$C_{\ell}$ at low $\ell$. In figure~\ref{fig:variance}, we show the
size of these variations as a function of $\ell$, which matches with
the expectation that the variance is entirely due to cosmic variance.
In the rest of the paper, we will simply treat our sensitivity as
cosmic variance limited.

\subsection{Method comparison and limitations}\label{sec:limitations}

In this section, we have shown two methods that we used to compute the two
point function of the polarization rotation angle. The two methods agree
very well at small values of $\theta$ but differ at large angular
scales (see figure~\ref{fig:twopointcompare}). This arises from the
fact that the analytical analysis does not take into account the long
strings present in the current epoch.  These long strings will
contribute to correlations at a very large angular scale and small
$\ell$ ($\ell \sim 1$). The toy simulation captures the large dipole
contribution from these strings stretching across the sky. However, as
shown in figure~\ref{fig:variance}, at small $\ell$, our prediction is
cosmic variance limited. 
The analytical method might also overestimate the effect coming from
the string loops that are either very close to the surface where CMB
photons are emitted or very close to the detector. For these string
loops, treating the CMB photons as originating from and ending at
infinitely far away leads to an over-estimate of the total phase
accumulated by the CMB photons, which might also lead to a difference
between the analytical calculation and numerical simulation.

The good agreement of the two methods at small angular scales is
encouraging since treating strings as a circular loop versus
infinitely long straight lines are two extremely different ways of
modeling the shape of strings in our universe. Getting comparable
results for the two cases suggest that the uncertainty coming from the
evolution of the string network should be small. However, precise
predictions of polarization rotation from an axion string
network extracted from a dedicated numerical simulation will provide
valuable information about the evolution of the network
as the universe transitions from radiation domination, to matter
domination, to CC domination today, and will be crucial for confirming
the nature of the signal in light of an observation. On the other
hand, an experimental measurement of strings will give us information
about the total
string length per Hubble as well as the domain wall number $N_{\rm
DW}$.  This information would be very useful for clarifying the
evolution of the string network or string-domain wall network, in
particular the deviation from a scaling solution, in more general
scenarios.\footnote{We thank
Giovanni Villadoro for inspiring discussions on this subject at GGI.} 

On the CMB side, we have not taken into account two issues that can
make the analysis more involved.  First, the CMB photons are not
emitted all at the same redshift~\cite{Hu:1997hv}. CMB photons emitted
from the same angular direction at different $z$ can have slightly
different polarization rotation angles.  The effect is largest for
the string closest to the surface of last scattering, which will cause
a difference in the polarization rotation angle of $\Delta \Phi \propto \xi
\Delta z/z$ where $\Delta z \sim 200$ is the width of the surface of
last scattering.  We can estimate the change in the two-point function
shown in equation~\eqref{eq: self 2 point} by taking one of the many strings
and giving it a rotation $\Phi = \anomaly \alphaem (1 - \xi \Delta z/z )$, as 
opposed to the usual $\anomaly \alphaem$, to arrive at
\begin{equation} \label{Eq: diffuse axion}
\frac{\Delta \langle \Phi(0,0)^2 \rangle}{\langle \Phi(0,0)^2 \rangle} \sim \frac{\Delta z}{z_{\rm CMB}} \frac{1}{\log \left(\eta_0/\eta_{\rm CMB} \right)} \sim 1\%
\end{equation}
so that this is not expected to be a significant effect.
Secondly, the polarization rotation angle power spectrum receives some additional contribution from the period of reionization at low $\ell$. These new contributions to the CMB polarization will only be affected by the large string loops present at late time, and will as a result, have a different rotation angle as compared to the polarization generated around the time of recombination. The contributions to the CMB polarization during the epoch of reionization is significant at low $\ell$ (e.g. it is dominant for $\ell \lesssim 10$ for $E$-mode polarization).  A robust prediction of the correlation functions at low $\ell$ will require a dedicated simulation. However, we do not expect it to significantly change the qualitative features since the polarization generated around reionization still goes through $\mathcal{O}(1)$ of the total string loops. In fact, the ratio of the polarization rotation angle of the polarization generated during reionization (lower-$\ell$) and recombination (higher-$\ell$) can be an interesting cross check in light of a discovery. We leave these issues to a more dedicated analysis in the future. 

When axions are massive compared to $H_0$, the strings and domain walls have a very different evolution history, as outlined in section~\ref{Sec: string basics}. However, the observable signal does not change significantly in the case where $N_{\rm DW} \neq 1$. In fact, an axion mass that is larger than $H_{\rm CMB}$ will reduce many of the uncertainties we have coming from the issues mentioned before, since the formation of the axion domain wall will shrink the region of space where the axion field is varying and as a result, reduce the fraction of photons that are emitted in the region where the axion field is changing.  
This would be especially helpful if we use the CMB to determine the value of the quantized polarization rotation angle from a single string.  
As mentioned before, if $N_{\rm DW} = 1$, the string network will disappear when the domain walls form and the energy density stored in the string-domain wall network is converted to axion radiation. 

Another limitation of our simple analysis comes from the ambient
axion density emitted by the string (and domain wall) network. The
ambient density of axions emitted by this network have properties
similar to those coming from a misalignment mechanism, which was
studied in detail in~\cite{Fedderke:2019ajk}.  
The two effects of the axion radiation are a rotation of the
polarization due to the differences in the axion field value from the time
of CMB to today, and a reduction of the $E$-mode polarization due to
smearing by the differing axion field values during the time of the
emission of CMB photons.
The smearing effect is maximal when the frequency of oscillation of
the axion is fast compared to the duration over which the CMB is
emitted.  As the spectrum of axions emitted by the string network is
not reliably known yet, we estimate the strongest that this bound
could be by assuming that all of the axion radiation is oscillating
fast and at the same frequency.  The constraint in our normalization
is that
\begin{equation}
\frac{\anomaly \alphaem}{\pi}\frac{a}{f} \leq 0.15 
\label{eq:axion-radiation}
\end{equation}
coming from the Planck polarization
measurement~\cite{Fedderke:2019ajk}.  Taking the energy in axion
radiation to be the same as in the strings, we find that $\anomaly\lesssim
\mathcal{O}(10)$.
Unlike effects coming from strings, the polarization rotation coming
from the axion radiation is not enhanced by square-root of the
number of strings in between CMB and today ($\sqrt{N_s}$).  As a
result, we do not expect them to cause any qualitative change to the
correlation functions we predict. 
However, the ambient background of axions may act as noise and inhibit
our ability to identify a single string and measure the quantized
polarization rotation angle around the string.

\section{CMB Observables} \label{Sec: CMB}

In this section, we present how the polarization rotation coming from
string networks map onto CMB observables.  We first review how the
polarization rotation angle power spectrum is calculated from the CMB
maps and compare current constraints and forecasts with the
predictions from our simulations described in section~\ref{Sec: pol}.
Afterwards, we discuss features of our
signal that would allow it to be separated from backgrounds.  The most
interesting feature of string networks are the discontinuities in the
polarization direction that appear in the sky. Throughout this
section, we will use capital letter $L$ and $M$ when describing the
polarization rotation angle $\Phi$ specifically and small letters $l$ and
$m$ when describing the CMB E- and B-modes specifically. We will keep
using the letter $\ell$ for general discussion.

\subsection{Polarization rotation angle power spectrum}

We denote the birefringence rotation as $\Phi(\hat{\gamma})$, as before.
The cosmic birefringence leaves the temperature field and the size of
the polarization unchanged and
rotates the polarization angle. This corresponds to 
changing the Stokes parameters as follows,
\begin{align}
  p(\hat{\gamma})
  &=Q(\hat{\gamma}) \pm i U(\hat{\gamma})
  =
  (\tilde{Q}(\hat{\gamma}) \pm i \tilde{U}(\hat{\gamma}))
  \exp(\pm 2i \Phi(\hat{\gamma}))
  \label{eq:stokes}
\end{align}
where $\tilde{Q}, \tilde{U}$ are the Stokes parameters in the absence
of birefringence. 
Since $U$ is parity-odd and $Q$ is parity-even, this rotation breaks
parity.
In terms of $E$- and $B$- mode decomposition,
\begin{align}
  p(\hat{\gamma})
  &=
  \sum_{lm} (E_{lm} + i B_{lm}) _2Y_{lm}(\hat{\gamma})
  \label{eq:EBmodes}
\end{align}
where $_2Y_{lm}$ are the spin-2 spherical harmonics. The polarization
rotation converts E-modes into B-modes in a $\ell$ dependent way. Since the
primordial $B$-modes have been constrained to be far smaller than the
$E$-modes, the dominant signal comes from generation of $B$-modes~\cite{Contreras:2017sgi,Gluscevic_2012},
\begin{align}
  B_{lm}
  &=
  2\sum_{LM}
  \sum_{l'm'}
  \Phi_{LM}
  E_{l'm'}
  \Xi^{LM}_{lml'm'}
  H^L_{ll'}
  \label{eq:Blm}
\end{align}
with $\Xi$ and $H$ are associated with Wigner-3$j$ symbols,
\begin{align}
  \Xi^{LM}_{lml'm'}
  &\equiv 
  \frac{(-1)^m}{4\pi}
  \sqrt{(2l+1)(2l'+1)(2L+1)}
  \left(
    \begin{array}{ccc}
      l & L & l' \\
      m & M & m'
    \end{array}
  \right)
\nonumber\\
  H^L_{ll'}
  &\equiv 
  \left(
    \begin{array}{ccc}
      l & L & l' \\
      2 & 0 & -2
    \end{array}
  \right)
  \label{eq:Wigner3j}
\end{align}
and we have expanded the polarization rotation angle in spherical harmonics,
$\Phi(\hat{\gamma}) = \sum_{LM} Y_{LM}(\hat{\gamma}) \Phi_{LM}$. 
The summation is restricted to even values of $L+l+l'$. This reflects
the unique parity violating nature of birefringence. For example,
$B$-modes generated by
lensing $E$-modes only get contributions from the odd values of the
sum, and thus can be distinguished from our signal.

This can be used to build an
estimator $\hat{\Phi}_{LM}$ for the polarization
rotation angle~\cite{Gluscevic:2009xxx,Gluscevic_2012}, for example
\begin{align}
  [\hat{\Phi}^{E^iB^j}_{LM}]_{ll'}
  &=
  \frac{2\pi}{(2l+1)(2l'+1)C_l^{EE} H_{ll'}^L}
  \sum_{mm'} B^i_{lm} E^{j*}_{l'm'} \Xi^{LM}_{lml'm'}
  \,.
  \label{estimator}
\end{align}
The minimum variance estimator combines estimates from all such channels,
$\{E^i B^j, B^iE^j, T^iB^j, B^i T^j\}$, where $i,j$ index channels
for a given map.

\begin{figure}[tp]
  \centering
  \includegraphics[width=0.85\textwidth]{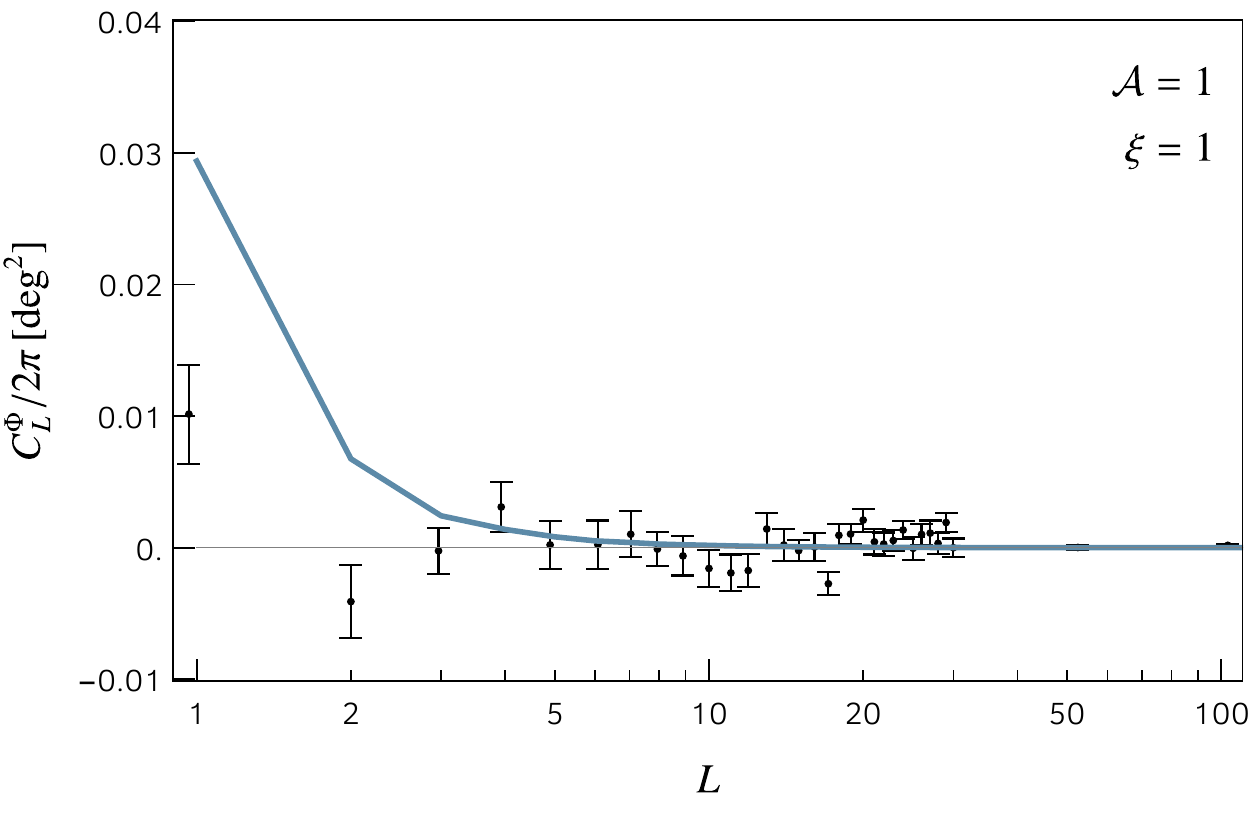}
  \\
  \includegraphics[width=0.85\textwidth]{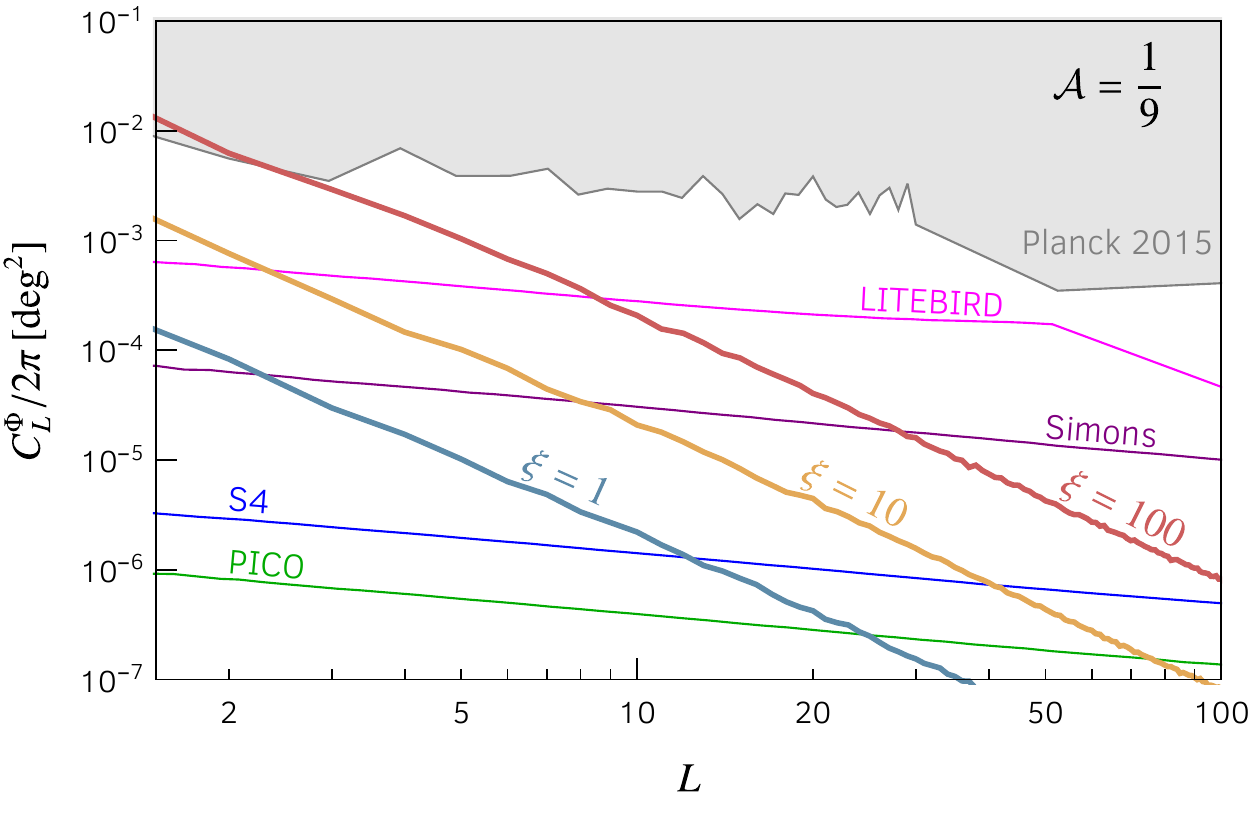}
  \caption{Power spectrum of the polarization rotation angle. Top:
  Constraints on the power spectrum derived from Planck 2015
polarization data (data points)~\cite{Contreras:2017sgi}. The blue
curve shows the power spectrum for $\xi =
\anomaly = 1$, which is of the order of the current sensitivity.
Bottom: Forecasts for the power spectrum measurements in
future experiments~\cite{Pogosian:2019jbt}. The grey shaded region is
excluded by Planck while the thin magenta, purple, blue and green
lines show the sensitivity of future CMB experiments. The thick blue, yellow
and red solid lines show the prediction for $\xi = 1,\,10,\,100$ and
$\anomaly = 1/9$. For both plots we have taken a mean over cosmic
variance for our prediction. The power spectra can be scaled for other
values of $\xi$ and $\anomaly$ as  $C_L^\Phi \propto \xi \anomaly^2$.} 
\label{fig:data}
\end{figure}

The variance in the estimator is given as,
\begin{align}
  \langle \hat{\Phi}_{LM} \hat{\Phi}_{L'M'} \rangle
  &=
  \delta_{LL'}\delta_{MM'} (C_L^\Phi + \sigma^2_{\Phi,L})
  \,,
  \label{eq:twopointmeasured}
\end{align}
where $\sigma^2_{\Phi,L}$ is the combined variance of the
estimators, and $C_L^\Phi$ is the power spectrum of the polarization
rotation angle due to birefringence. The variance $\sigma^2_{\Phi L}$
includes contributions from
the beam systematics, detector noise, galactic and atmospheric
foregrounds and weak lensing contribution.
This variance can then be used to forecast sensitivity to $C_L^\Phi$,
\begin{align}
  \sigma_{C_L^\Phi}
  &=
  \frac{\sigma^2_{\Phi,L}}{f_{\rm sky} (2L+1)/2}
  \,.
  \label{eq:sigmaCL}
\end{align}
The details of the sky coverage $f_{\rm sky}$, delensing fraction,
beam parameters
and sensitivity used for different experiments in
figure~\ref{fig:data} can be found in~\cite{Pogosian:2019jbt}.  The
current CMB constraints are sensitive to a constant birefringence
rotation of $0.5^{\circ}$ at $1\sigma$ level, which is roughly of the
expected size $\Phi \simeq \alphaem$ for $\xi = 1$ and $\anomaly = 1$. 
In figure~\ref{fig:data} (top) we plot the $C_L^\Phi$ values obtained
in our simulation (for $\xi=1$, $\anomaly=1$) against the $C_L^\Phi$
constraints derived in~\cite{Contreras:2017sgi} from Planck. We also
plot future sensitivity ($\sigma_{C_L^\phi}$) for various future
experiments~\cite{Pogosian:2019jbt} (bottom). The orthogonality with
the lensing signal implies that the optimal sensitivity is obtained
for measurements around the peak of the E-mode spectrum, $500< \ell <
3000$. Higher resolution does not lead to better sensitivity to the
rotation power spectrum since the polarization signal is smaller at
smaller scales. However, as we will see below, higher resolution can
help resolve the strings directly in position space.

It is intriguing to note that there is an excess dipole signal 
in the data. If this is a signal of a string, it may be
possible to find further evidence for it in the newly released Planck
2018 data. Future experiments will have improvement in sensitivity by
orders of magnitude, which can
definitively probe a signal coming from $\xi, \anomaly\sim 1$ string
networks.

\subsection{Distinguishing strings from other sources of B-modes}
In this section, we briefly address the distinguishability of our
signal from other sources of B-mode polarization. Our signal, as shown
in figure~\ref{fig:data}, will have a particular $\ell$-dependence,
which can potentially be distinguished from B-modes coming from other
sources. Moreover, we expect our effect to have a distinct structure
when we consider higher point correlation functions. In particular, we
expect the 3-point correlation function and all odd correlation
functions are zero at leading order, and all the even $N$-point
correlation functions to have a scaling with $\xi$ and $\anomaly
\alphaem$ as $\xi (\anomaly \alphaem)^N$.  Furthermore, there are two
other main features of our signal at the level of two point
correlation functions of $\Phi$ in the CMB, parity violation and
frequency independence.

\paragraph{\underline{Lensed B-modes}} The lensing of $E$-modes produces correlated $B$
modes in the CMB spectrum. This leads to parity conserving
correlators, and hence are distinguishable to the axion induced
birefringence signal.

\paragraph{\underline{Primordial Gravitational Waves}}
Primordial gravitational waves induced by high scale inflation are a
prime target for CMB experiments. The gravitational waves induce a
$B$-mode spectrum, but not a parity violating $EB$ correlators in
most inflation
models~\cite{1985SvA....29..607P,Seljak:1996gy,
Kamionkowski:1996zd,Seljak:1996ti}.
In specific models~\cite{Anber:2012du,Adshead:2013qp,Thorne:2017jft},
the spectrum of gravitational waves induced  can be chiral which leads to
parity-violating fluctuations in the CMB. These models come with a
rich set of associated signals such as non-gaussianities, and we
expect that detailed properties of the power spectrum and higher-point
functions will readily distinguish this signal from axion strings.

\paragraph{\underline{Faraday rotation}}
Primordial magnetic fields generate a Faraday rotation of the CMB
polarization, converting E modes to B modes. Such a primordial
magnetic field is motivated by observation of $\mu$G galactic magnetic fields
whose origin is not well-understood. The spectrum of the polarization
rotation angle induced by these magnetic fields depends on their power
spectrum. Faraday rotation has a distinctive frequency dependence,
$\Phi \sim \nu^{-2}$, and hence it will be very easily distinguished
from an axion string induced rotation. 

\paragraph{\underline{Lorentz violation}}
Cosmic Birefringence was in fact first proposed~\cite{Carroll:1989vb}
in the context of measuring Lorentz violating terms in the SM. In the
SM extension formalism, the $d=3$ terms associated with
a Chern-Simons-like coupling is physically equivalent to an axion
background with a uniform gradient. Higher dimensional terms produce
frequency dependent terms. Therefore, it will be relatively easy to
disentangle effects of such terms from an axion string network.

We see that even at the statistical level, it will likely be possible
to disentangle our signal from other potential sources of parity-odd
$B$-modes in the CMB. However, the spatial morphology of the signal
from strings, with a $\sim \anomaly \alphaem$ jump across the strings,
is a smoking gun feature which sets apart the axion string signal.
This is the signal we now turn to.

\subsection{Edge detection}
Cosmic strings are distinctive features in position space, so
analyzing the signal in Fourier space may not be the optimal
strategy to study the properties of the strings. For this reason, it would be interesting to look for the
strings directly in position space. For the case of gravitational
effects of cosmic strings, edge detection algorithms for the CMB
temperature maps~\cite{Stewart:2008zq} have been proposed.

\begin{figure}[tp]
  \centering
  \includegraphics[width=0.3\textwidth]{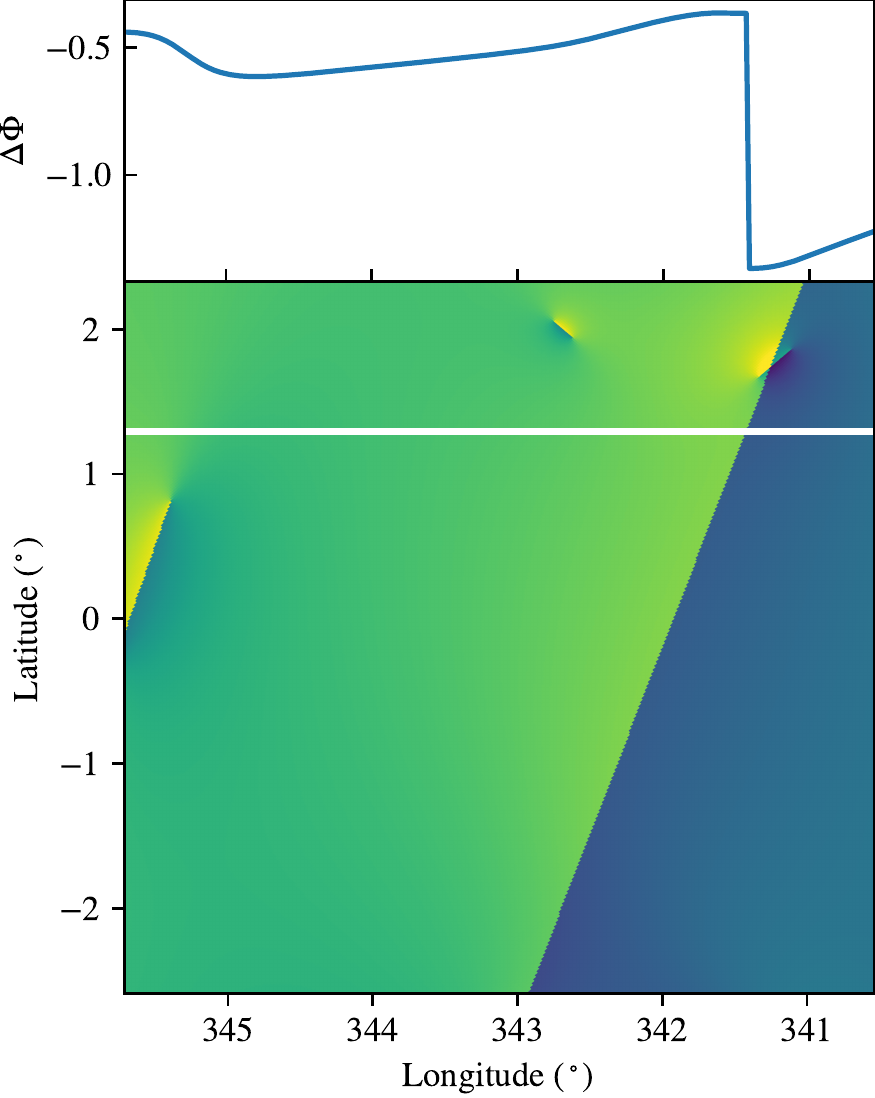}
  \quad
  \includegraphics[width=0.3\textwidth]{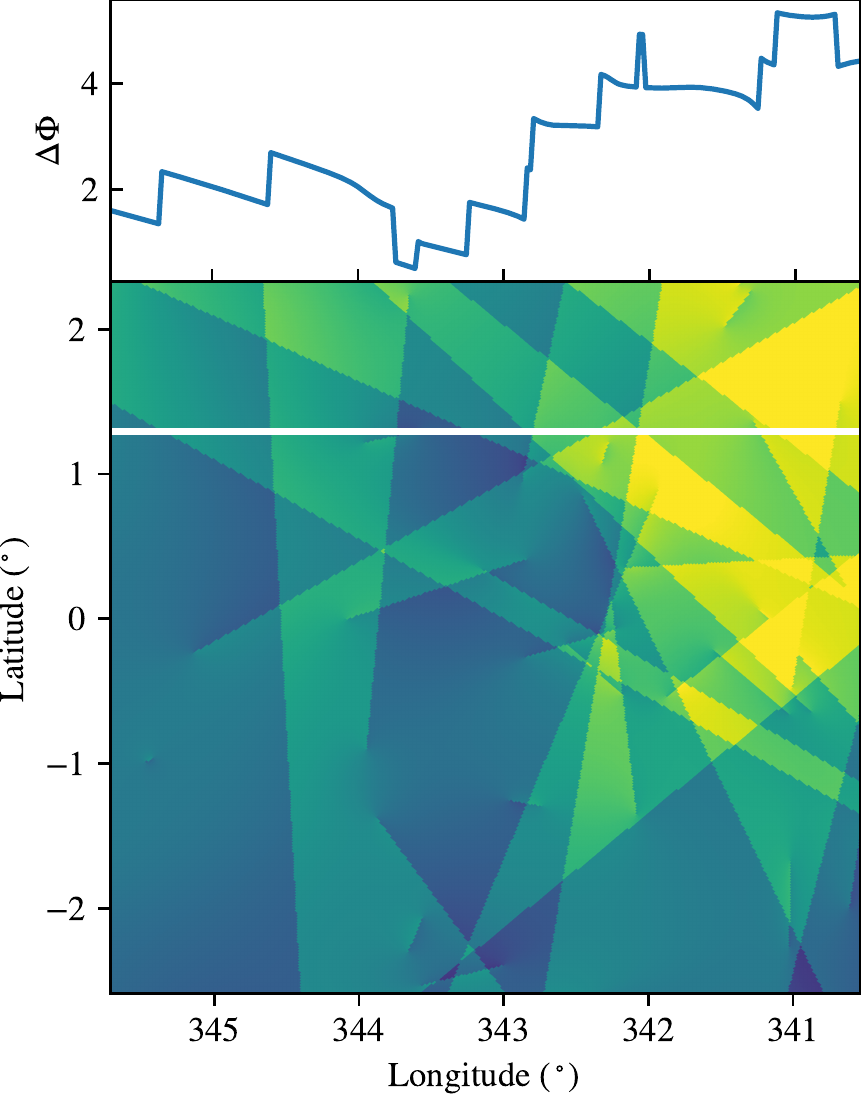}
  \quad
  \includegraphics[width=0.3\textwidth]{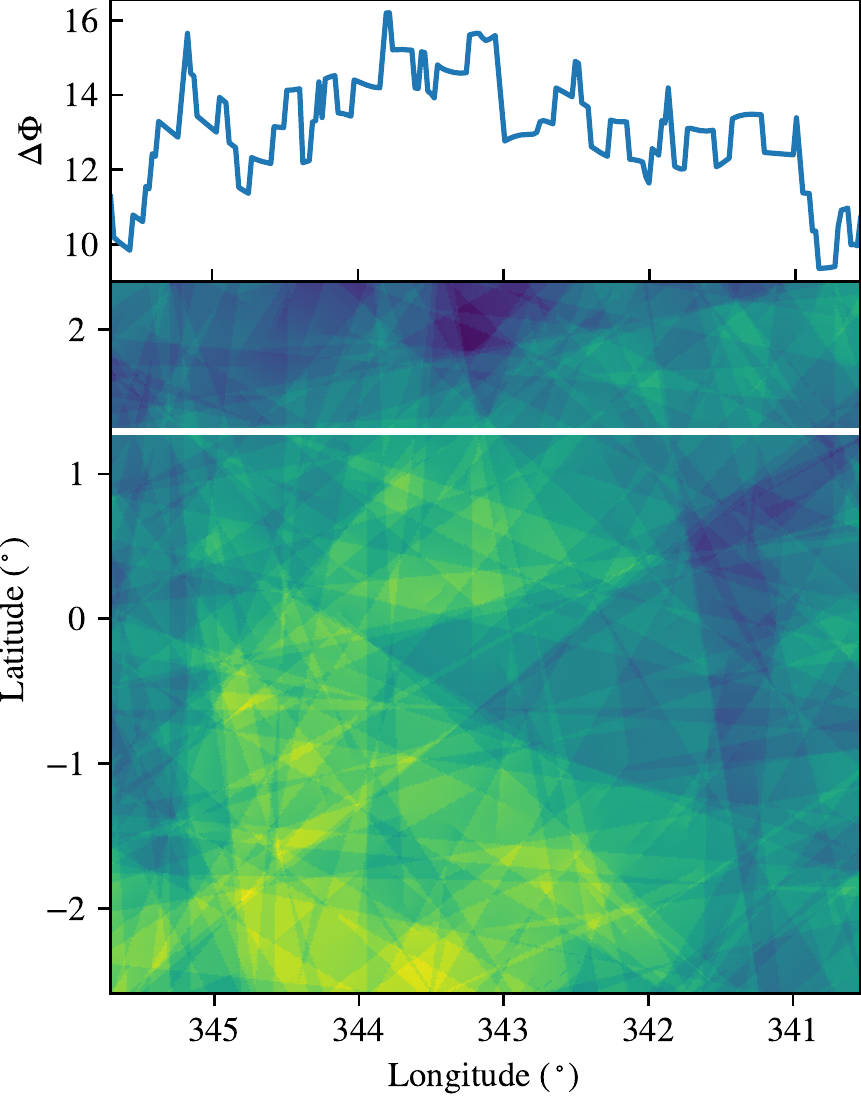}
  \caption{Zoomed-in skymaps at the truth-level for $\xi=1,10,100$
    showing the white
    framed areas in the respective skymaps in figure~\ref{fig:skymap}
    (the color scale is modified for visual clarity). 
    The presence of sharp discontinuities is apparent in these images
    of broken light.
    The
    discontinuity is quantized and is equal to 1. The white
    stripe shows a slice through the map, and the value of $\Delta
    \Phi$ (normalized by $1/(\anomaly\alphaem)$) is shown in the top panels
    where the quantized jump is apparent.
  }
  \label{fig:disc}
\end{figure}

The polarization rotation angle of the photons has a discontinuity across
the axion string.
A similar edge detection algorithm
can be developed to detect this discontinuity in polarization maps.
The value of the discontinuity is an extremely interesting quantity to
measure. The two path across the string differ by a path that loops
around the string, and hence measures the topological charge of the
string (see section~\ref{sec:implications}).

In figure~\ref{fig:disc}, we show a zoomed-in
patch of our simulation, where the discontinuity is apparent. To
develop this edge detection algorithm for polarization maps will be an extremely interesting
future direction. The size of the discontinuity contains a wealth of
information, as we have said, so to extract that would be important. Edge detection can potentially be used to look for a single string in the entire universe, or strings that are even beyond our horizon~\cite{Kaplan:2008ss}, both corresponding to cases where the $U(1)$ symmetry is broken before inflation.

The two main limiting factors on edge detection are the angular
resolution of the experiment and the precision of polarization
measurement. Ground-based observatories like The Atacama Cosmology
Telescope (ACTPol)~\cite{Thornton:2016wjq}, South Pole
Telescope~\cite{2011PASP..123..568C},
POLARBEAR~\cite{2012SPIE.8452E..1CK}, and future Simons
Observatory~\cite{Ade:2018sbj} will have the ability to measure
polarizations at arcmin angular scales. This corresponds to a
sensitivity in the range $\xi\lesssim 100$ such that the angular
separation between strings is larger than the angular resolution. The
Simons Observatory~\cite{Ade:2018sbj} has a noise-level of $\sim {\mu
{\rm K}}$-arcmin, and the ability to measure polarization rotations to
$\mathcal{O}({\rm deg})$. Improvements of both the angular resolution
and noise-level would improve the sensitivity to axion strings with
edge detection~\cite{2018SPIE10700E..5XP,Sehgal:2019ewc}. In
figure~\ref{fig:money}, we show a very rough estimate of the
sensitivity to axion strings using edge detection with current
technology and leave a dedicated analysis to future work. How the sensitivity scales with $\xi$ is not very clear
because, on one hand, having more strings requires better angular
resolution and smaller noise at small angular scales, while on the
other hand, having more strings also increases the statistics. Edge
detection in 2d images is a prime application for machine learning
algorithms.
Since our signal lives precisely in this space, it may be an ideal
candidate to apply some of these deep learning technologies to. It is 
a very interesting future direction to study edge detection techniques
to find individual strings in the presence of complications such as 
reconstruction noise.

\section{Other observational signature of axion strings} \label{Sec: other}

\begin{figure}[tp]
  \centering
  \includegraphics[width=0.85\textwidth]{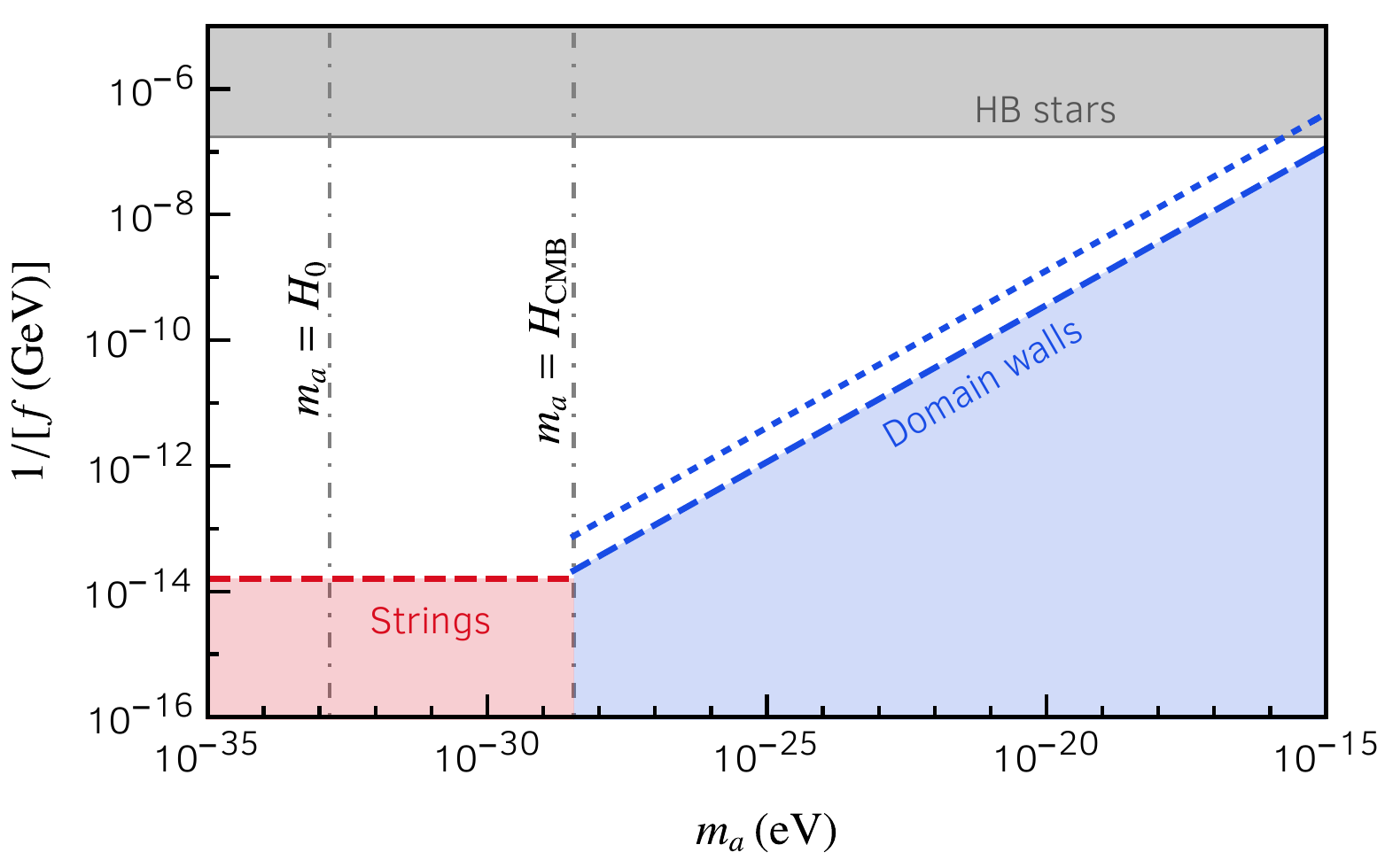}
  \caption{Parameter space for polarization rotation and gravitational
    wave signatures. The gray shaded region is excluded by cooling
    bounds coming from Horizontal Branch stars~\cite{Ayala:2014pea}
    (assuming $\mathcal{A}=1$).  The red shaded region is excluded by
    CMB measurements of strings coming from their gravitational
    effects~\cite{Charnock:2016nzm}.  The blue region is excluded when
    $N_{\rm DW} \ne 1$ due to the temperature fluctuations induced by
    the gravitational potential of the domain wall~\cite{Sousa:2015cqa}.
    Dashed and dotted lines correspond to $\xi=10$ and 100
    respectively. The figure is cut off at large $f$, as for $f
    \gtrsim 10^{16}\,{\rm GeV}$, the universe can not reheat to high
    enough temperature so that strings might not be produced. Our
    sensitivity applies to $\xi\geq 1$ in all of the allowed parameter
  space in this plot.}
  \label{fig:parameter-space}
\end{figure}

In this section, we discuss some of the other signatures of an axion string due to its coupling with the photon. We also discuss the signatures that come purely from the gravitational coupling of the axion strings and domain walls. These include their gravitational effect on the CMB as well as gravitational wave emission during the string network evolution. We collect all these constraints and future sensitivities in figure~\ref{fig:parameter-space}.
\subsection{Quasar lensing}

The axion string network can also be looked for with distant
gravitationally lensed polarized light sources. A particular example
of such a source is a quasar. Recently, many strongly gravitationally
lensed quasars have been found and carefully measured to extract
valuable information about the expansion history of our universe by
the H0LiCOW collaboration~\cite{Suyu:2016qxx}. These strongly lensed
quasars, with multiple images and image separations as large as $\beta
\sim 20''$~\cite{Inada:2003vc} (see \cite{glqdatabase} for a list of
205 known gravitational lensed quasars), can also provide very unique
information about cosmic axion strings.

\begin{figure}[tp]
  \centering
  \includegraphics[width=0.65\textwidth]{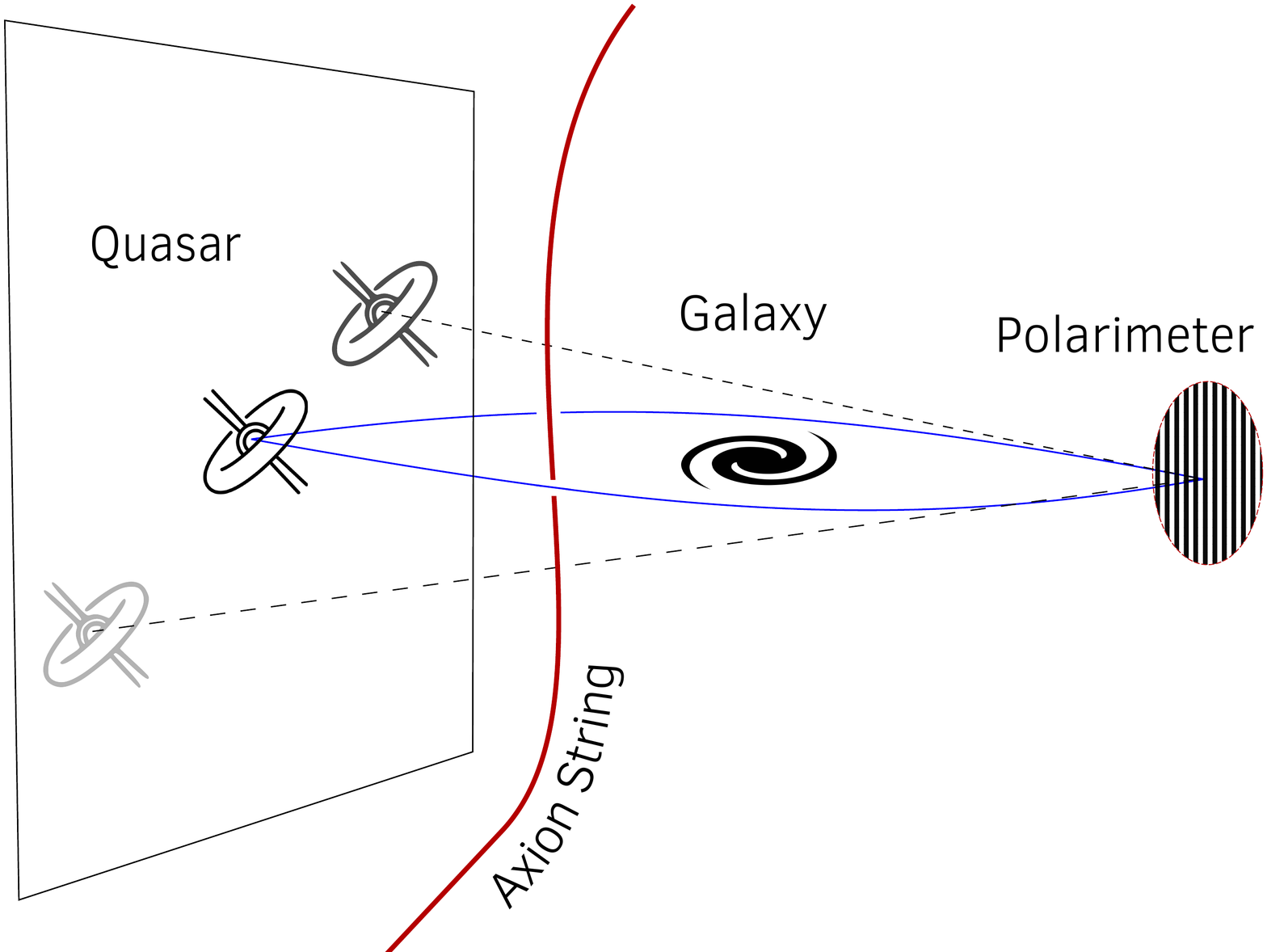}
  \caption{A quasar at a cosmological distance is gravitationally
  lensed by a galaxy or galaxy cluster in between the quasar and the
observer. The different photon trajectories (blue lines) from the
quasar to the observer, corresponding to different quasar images, form
a closed loop around the gravitational lens. An axion string (red
line) can pass through this closed loop of photon trajectories,
leading to a differential polarization rotation angle between the two
photon trajectories (quasar images). }
  \label{fig:lensing}
\end{figure}

As shown in figure~\ref{fig:lensing}, a lensed quasar can have multiple images with angular separation $\beta$.  The area enclosed by the two photon trajectories is roughly
\begin{equation}
A_{\rm enc} \simeq D_L D_S \beta/2,
\end{equation}
where $D_S$ and $D_L$ is the distance to the source and the lens,
respectively. If the enclosed area is pierced by an axion string, the
two observed image will have a relative rotation between the
polarization of their images.  As before, this rotation will be
quantized.  In the case of the CMB, the approximately quantized nature
of the signal stems from the fact that for most string loops are at an
intermediate redshift, and the CMB photon can be treated as either
traversing through a loop or not.  In contrast, the quantized nature
of the lensed quasar signal comes from the fact that the two photon
trajectories combine to actually form a closed curve. This makes
gravitational lensed quasars, and other strongly lensed distant
objects unique systems to test the existence of these axion strings.
The probability for an axion string to traverse this area is
\begin{equation}
p \simeq \xi A_{\rm enc} H_0^2 \approx 10^{-3} \frac{\xi}{100}\frac{\beta}{10''},
\end{equation}
for each lensed quasar that is at a cosmological distance.

Measurement of the polarization of distant quasars have been made in the optical~\cite{1990ApJ...354..124I,Sluse:2005bg,1998A&A...340..371H} and the radio frequency range~\cite{Taylor_1998,Taylor:1999ms}.  The polarization uncertainties in this frequency range are at the degree level, mainly coming from calibration, which can potentially be improved when measuring relative polarization of two quasar images. 
Our signal is a quantized polarization rotation at the \% level or smaller.
Current easily accessible technology for measuring the polarization of optical frequency photons is around the $\sim 10^{-5}$ level~\cite{2017MNRAS.465.1601B}.
Thus, future observations of lensed polarized quasars can complement
the search with the cosmic microwave background, and have the unique
ability to test the quantized nature of the observed signal.
It will be very interesting to perform
detailed projections for searches with lensed quasars. We leave this
for future work.

The distinguishability of our signal from background relies on two important properties of the signal coming from a topological defect. The topological nature of the signal ensures that the polarization rotation angle is quantized. As a result, different pairs of lensed quasar images, as well as different lensed quasar systems, will have the same relative polarization rotation up to an integer multiple\footnote{Given that the axion strings should be very rare in our universe, multiple strings traversing through the same lens is very unlikely.}.  The topological nature of the signal also ensures that the polarization rotation angle is frequency independent, which helps when distinguishing our effect from that of Faraday rotation along different trajectories. The axion strings we can look for with quasar lensing will have lengths that are comparable to the size of the universe today and will also leave an imprint on the cosmic microwave background, providing another unique cross check in the case of a discovery. 

\subsection{Gravitational and gravitational wave signatures}

The strings and domain walls can also be looked for through their gravitational effect on the large scale structures of the universe. The measurement of the detailed features of, in particular, the cosmic microwave background spectrum place a constraint on the amount of density inhomogeneities that can be generated by cosmic strings~\cite{Bevis:2007gh,Battye:2010xz} and domain walls. These considerations place an upper bound on the energy per unit length of the string $\mu$~\cite{Charnock:2016nzm},
\begin{equation}
G\mu \lesssim 1.1 \times 10^{-7},
\end{equation}
corresponding to an axion decay constant $f$ no higher than $\sim 10^{14} \,{\rm GeV}$. Such a constraint on the axion decay constant is only mildly dependent on the deviation from scaling solution (see the discussion around equation~\eqref{Eq: energy} in section~\ref{Sec: string basics}).

Similar constraints can be placed on the tension of domain walls. The energy density in the form of domains walls in a frustrated string-domain wall network approach a scaling limit~\cite{Hiramatsu:2012sc,Hiramatsu:2013qaa} for $N_{\rm DW} \neq 1$,
\begin{equation}
  \rho_{\rm wall} \approx N_{\rm DW} \sigma H,
\end{equation}
where $\sigma \sim 8 m_a f^2$ is the tension of the domain wall~\cite{Saikawa:2017hiv}. Measurements of the cosmic microwave background place a constraint on the tension of the domain walls~\cite{Sousa:2015cqa}
\begin{equation}
N_{\rm DW} G{\sigma} L_0 \leq 5.6\times 10^{-6},
\end{equation}
where $L_0 \approx 1/H_0$ is the characteristic size of the domain
wall today. This is an updated version 
of the famous Zel'dovich-Kobzarev-Okun bound of
${\sigma} \lesssim \mathcal{O}({\rm MeV^3})$~\cite{Zeldovich:1974uw}.

Similar to the case of cosmic strings, the frustrated string-domain
wall network can also have densities that deviate logarithmically from
the scaling solution when the density of strings and domain walls are
comparable.  It is unclear if such deviation would persist when the
domain wall energy density dominates that of the strings at late
time. In figure~\ref{fig:parameter-space}, we show the constraints
from gravitational measurement of the string-domain wall network
assuming that the string-domain wall network deviates from a scaling
solution logarithmically, similar to that of a string network.

The string network and the string-domain wall network emit
gravitational waves as they evolve in time.  The energy density and
spectrum of gravitational waves from a string network have been
studied extensively in the
literature~\cite{Vilenkin:1981bx,Vachaspati:1984gt,Battye:1994au}
(see~\cite{Cui:2018rwi} for more details).  Future gravitational wave
detectors can probe an axion decay constant as small as $\sim 10^{15}
\,{\rm GeV}$ at intermediate frequencies (LISA) and  $\sim 10^{14}
\,{\rm GeV}$ at low frequencies (SKA)~\cite{Chang:2019mza}.  The
string-domain wall network can produce gravitational waves as it
redshifts as well as when domain walls collide or annihilate.
The gravitational wave signals associated with these domain walls are observable only in regions of parameter space where the domain walls annihilate at a very early time~\cite{Saikawa:2017hiv} and occupy a range of parameter space very different from what we are probing. 

\section{Conclusion} \label{Sec: conclusion}

\begin{figure}[tp]
\centering
\includegraphics[width=0.85\textwidth]{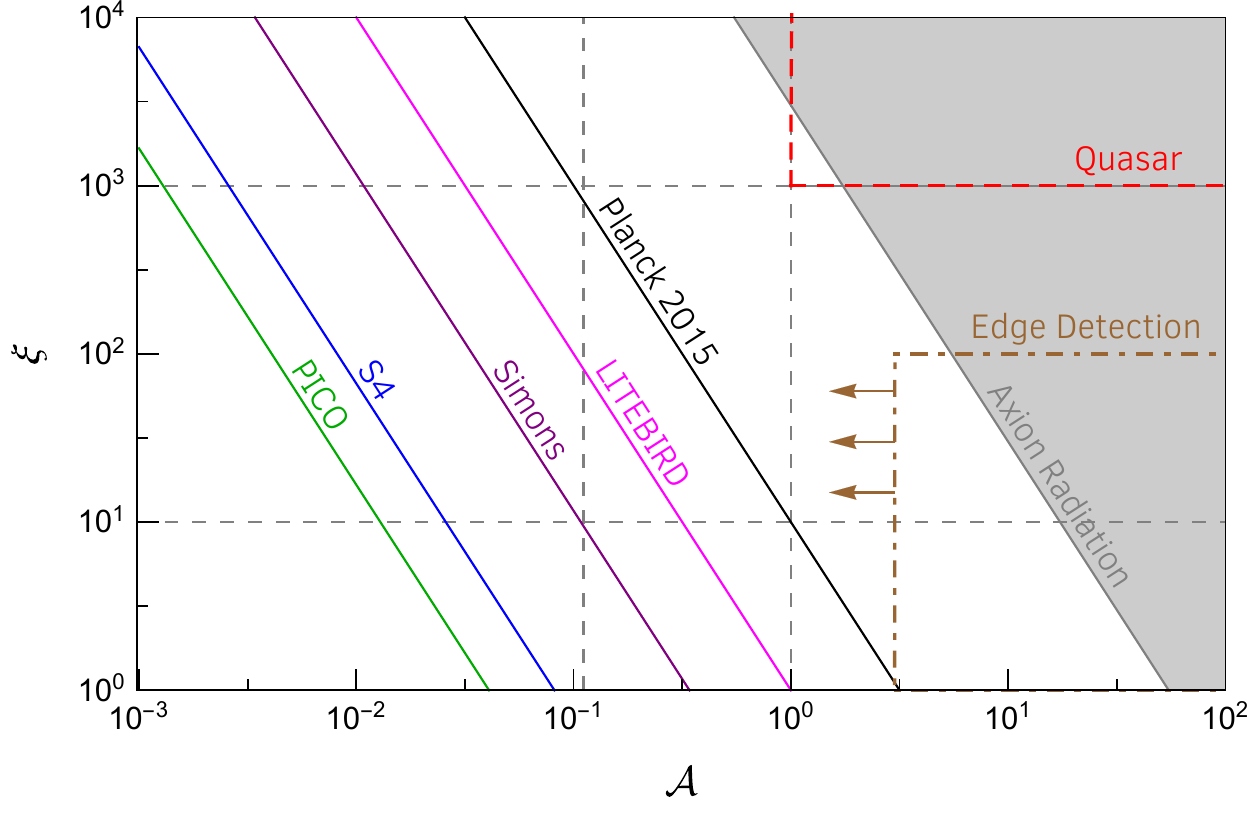}
\caption{We show the sensitivity of current and future experiments to
the two independent parameters of an axion string network with a
photon coupling; the number of axion string per Hubble volume ($\xi$)
and the strength of the axion string-photon coupling ($\anomaly$). The
CMB polarization correlation functions provide the best channel for
discovery (they put an upper bound on the combination
$\xi\anomaly^2$). The black solid line is the estimated current
sensitivity of
Planck to this effect while the magenta, purple, blue and green
lines show the sensitivity of future CMB experiments. We take the
sensitivity at $L=10$ from figure~\ref{fig:data} as an indicator. The
grey region
is the estimated constraint coming from the effects of axion radiation
emitted from the string network, see the discussion around
equation~\eqref{eq:axion-radiation}.  CMB experiments with good angular
resolution and lensed quasar systems provide an opportunity to exploit
the quantized nature of our signal. The brown dot-dashed line shows
the prospects for edge detection with experiments similar to the
Simons observatory (very qualitative). Increasing the accuracy of the
polarization rotation angle measurement improves the sensitivity to
smaller $\anomaly$ and improving the angular resolution improves the
sensitivity to higher $\xi$. The red dashed line shows the prospects
for lensed quasar systems (qualitative). Increasing the number of
lensed quasars improves the sensitivity to smaller $\xi$ while
improving the precision of the measurement of the relative
polarization of the quasars improves the sensitivity to smaller
$\anomaly$. The gray dashed gridlines mark the regions of parameter
space of particular theoretical interest.}\label{fig:money}
\end{figure}

String theory has reinforced the motivation for many phenomenological
paradigms at low energies, two of which are
cosmic strings~\cite{Witten:1985fp,
Polchinski:2004hb,Copeland:2003bj,Copeland:2004iv,Polchinski:2004ia}
and light bosonic states
(axiverse)~\cite{Arvanitaki:2009fg,Demirtas:2018akl}. In this paper,
we have shown that these are amusingly connected to each other: the
axion-photon
coupling offers a unique way to look for cosmic strings, while
cosmic axion strings also offer a unique way to look for axiverse
axions.

Due to the extended nature of cosmic axiverse strings and the
topological interaction between the strings and photon shown in
equation~\eqref{eq: best interaction}, measurement of cosmic axiverse strings
provide a unique opportunity to probe the UV dynamics of the theory.
Measurement of the UV anomaly coefficient $\anomaly$ can teach us
about fundamental physics such as the smallest unit of electric
charge.  Aside from being theoretically fascinating, cosmic axiverse
strings are also exciting phenomenologically.  The basic way of
searching for cosmic axiverse strings using photons is that the
polarization direction of linearly polarized light rotates by a
quantized amount when a photon circles a string.  If the UV theory
only has unit electrically charged particles, this quantized rotation
will be of order a percent.  If the UV theory is SM-like in its charge
assignments, then the
quantized rotation will be larger than $0.1\%$.

One of the most interesting ways in which this quantized rotation can
be measured is using the CMB.  The CMB provides a backlight which
shines upon all of the strings.  The difference in paths for photons
passing to the left and to the right of the string is a loop around
the string; thus the relative polarization rotation angle of the
photons is quantized.
The effect of strings on polarization can be searched for using a
standard angular decomposition (power spectrum) with the Planck
satellite and future CMB missions. What is perhaps more exciting is
the fact that the polarization map contains discontinuities at the
location of the strings, which can be looked for using edge detection
techniques with ground-based missions. Finding these strings in
position space and their associated discontinuities provides a direct
measurement of the anomaly coefficient induced by a single string.

In this paper, we have provided a description and rough estimate of
the observational features of a string network on the CMB (see
figure~\ref{fig:money}).  This analysis can be improved upon with real
axion string simulations that would capture more details of the nature
of the string network, and with a dedicated analysis of the CMB data.
Given the uncertainties and difficulties present in the simulations
due to simulating vastly different length scales, in the event of a
discovery one might even use data to gain insights for simulations
of axion strings and domain walls, which can be applied to the study
of other axion models such as the QCD axion. The combination of the
measurement of the CMB polarization spectrum and the discontinuities
coming from individual strings allows us to extract the value of $\xi$,
which will help with understanding the scaling solution and its
violation. 

Measurements of lensed quasar systems can offer complementary
information about the cosmic axiverse string network (see
figure~\ref{fig:money}). In particular, it offers a setup in our
universe where the photon one measures form a closed Aharonov-Bohm
like loop. As a result, we can use these lensed quasar systems to
measure the quantized phase without contamination coming from ambient
axion radiation.

The string axiverse has been a very exciting paradigm over the past ten
years. In this paper, we add a new tool to explore the axiverse. We
have shown that axion strings offer
a compelling way to look for the string axiverse that is free of the
assumption of these axions being the dark matter of our universe.
Cosmic axiverse strings provide novel untapped theoretical and
observational opportunities and it will be exciting to see where they
lead us.

\section*{Acknowledgement} 
The authors thank Martin Schmaltz for
collaboration during the early stages of the project. The authors
acknowledge Asimina Arvanitaki, Matthew Johnson, Amalia Madden,
Gustavo Marques-Tavares, Julian Mu\~noz, Lisa Randall and Matt Reece
for helpful discussions and valuable comments on the draft. The
authors also want to thank Masha Baryakhtar, Dagoberto Contreras,
Savas Dimopoulos, Michael Fedderke, Raphael Flauger, Andrei Frolov, Mathew Madhavacheril, Liam McAllister, Ken Olum,
Davide Racco,  Kendrick Smith, Matthew Strassler, Alex Vilenkin, Giovanni Villadoro for
useful conversations.  The authors acknowledge the KITP for its
hospitality during the inception of this project, supported partly by
National Science Foundation under Grant No. NSF PHY-1748958. JH would
like to express a special thanks to the GGI Institute for Theoretical
Physics for its hospitality and support.
PA is supported by NSF grants PHY-1620806 and PHY-1915071, the Chau Foundation
HS Chau postdoc support award, the Kavli Foundation grant Kavli Dream Team, and the
Moore Foundation Award 8342.
AH is supported in part by the NSF
under Grant No. PHY-1914480 and by the Maryland Center for Fundamental
Physics (MCFP).  Research at Perimeter Institute is supported by the
Government of Canada through Industry Canada and by the Province of
Ontario through the Ministry of Economic Development \& Innovation.

\bibliographystyle{JHEP}
\bibliography{strings}

\end{document}